\documentclass[twocolumn]{aastex62}
\usepackage{subfigure}
\usepackage{systeme}
\usepackage{longtable}
\usepackage{mathtools}
\usepackage{amsmath}

\DeclareMathOperator\erf{erf}

\shorttitle{MWA Phase II EoR Power Spectrum}
\shortauthors{Li et al.}

\begin{document}

\title{First Season MWA Phase II EoR Power Spectrum Results at Redshift 7}

\correspondingauthor{Jonathan Pober}
\email{Jonathan\_Pober@brown.edu}

\author{W.~Li}
\affiliation{Department of Physics, Brown University, Providence, RI 02912, USA}
\author{J.~C.~Pober}
\affiliation{Department of Physics, Brown University, Providence, RI 02912, USA}
\author{N.~Barry}
\affiliation{School of Physics, The University of Melbourne, Parkville, VIC 3010, Australia}
\affiliation{ARC  Centre  of  Excellence  for  All  Sky  Astrophysics  in  3  Dimensions  (ASTRO  3D)}
\author{B.~J.~Hazelton}
\affiliation{Department of Physics, University of Washington, Seattle, WA 98195, USA}
\affiliation{University of Washington, eScience Institute, Seattle, WA 98195, USA}
\author{M.~F.~Morales}
\affiliation{Department of Physics, University of Washington, Seattle, WA 98195, USA}
\affiliation{ARC  Centre  of  Excellence  for  All  Sky  Astrophysics  in  3  Dimensions  (ASTRO  3D)}
\author{C.~M.~Trott}
\affiliation{International Centre for Radio Astronomy Research (ICRAR), Curtin University, Bentley, WA 6102, Australia}
\affiliation{ARC  Centre  of  Excellence  for  All  Sky  Astrophysics  in  3  Dimensions  (ASTRO  3D)}
\author{A.~Lanman}
\affiliation{Department of Physics, Brown University, Providence, RI 02912, USA}
\author{M.~Wilensky}
\affiliation{Department of Physics, University of Washington, Seattle, WA 98195, USA}
\author{I.~Sullivan}
\affiliation{Department of Physics, University of Washington, Seattle, WA 98195, USA}
\author{A.~P.~Beardsley}
\affiliation{School of Earth and Space Exploration, Arizona  State University, Tempe, AZ 85287, USA}
\author{T.~Booler}
\affiliation{International Centre for Radio Astronomy Research (ICRAR), Curtin University, Bentley, WA 6102, Australia}
\author{J.~D.~Bowman}
\affiliation{School of Earth and Space Exploration, Arizona  State University, Tempe, AZ 85287, USA}
\author{R.~Byrne}
\affiliation{Department of Physics, University of Washington, Seattle, WA 98195, USA}
\author{B.~Crosse}
\affiliation{International Centre for Radio Astronomy Research (ICRAR), Curtin University, Bentley, WA 6102, Australia}
\author{D.~Emrich}
\affiliation{International Centre for Radio Astronomy Research (ICRAR), Curtin University, Bentley, WA 6102, Australia}
\author{T.~M.~O.~Franzen}
\affiliation{ASTRON, Netherlands Institute for Radio Astronomy, Oude Hoogeveensedijk 4, 7991 PD, Dwingeloo, The Netherlands}
\affiliation{International Centre for Radio Astronomy Research (ICRAR), Curtin University, Bentley, WA 6102, Australia}
\author{K.~Hasegawa}
\affiliation{Kumamoto University, Japan}
\author{L.~Horsley}
\affiliation{International Centre for Radio Astronomy Research (ICRAR), Curtin University, Bentley, WA 6102, Australia}
\author{M.~Johnston-Hollitt}
\affiliation{International Centre for Radio Astronomy Research (ICRAR), Curtin University, Bentley, WA 6102, Australia}
\author{D.~C.~Jacobs}
\affiliation{School of Earth and Space Exploration, Arizona  State University, Tempe, AZ 85287, USA}
\author{C.~H.~Jordan}
\affiliation{International Centre for Radio Astronomy Research (ICRAR), Curtin University, Bentley, WA 6102, Australia}
\affiliation{ARC  Centre  of  Excellence  for  All  Sky  Astrophysics  in  3  Dimensions  (ASTRO  3D)}
\author{R.~C.~Joseph}
\affiliation{International Centre for Radio Astronomy Research (ICRAR), Curtin University, Bentley, WA 6102, Australia}
\affiliation{ARC  Centre  of  Excellence  for  All  Sky  Astrophysics  in  3  Dimensions  (ASTRO  3D)}
\author{T.~Kaneuji}
\affiliation{Kumamoto University, Japan}
\author{D.~L.~Kaplan}
\affiliation{Department of Physics, University of Wisconsin--Milwaukee, Milwaukee, WI 53201, USA}
\author{D.~Kenney}
\affiliation{International Centre for Radio Astronomy Research (ICRAR), Curtin University, Bentley, WA 6102, Australia}
\author{K.~Kubota}
\affiliation{Kumamoto University, Japan}
\author{J.~Line}
\affiliation{International Centre for Radio Astronomy Research (ICRAR), Curtin University, Bentley, WA 6102, Australia}
\affiliation{ARC  Centre  of  Excellence  for  All  Sky  Astrophysics  in  3  Dimensions  (ASTRO  3D)}
\author{C.~Lynch}
\affiliation{International Centre for Radio Astronomy Research (ICRAR), Curtin University, Bentley, WA 6102, Australia}
\affiliation{ARC  Centre  of  Excellence  for  All  Sky  Astrophysics  in  3  Dimensions  (ASTRO  3D)}
\author{B.~McKinley}
\affiliation{International Centre for Radio Astronomy Research (ICRAR), Curtin University, Bentley, WA 6102, Australia}
\affiliation{ARC  Centre  of  Excellence  for  All  Sky  Astrophysics  in  3  Dimensions  (ASTRO  3D)}
\author{D.~A.~Mitchell}
\affiliation{CSIRO Astronomy and Space Science (CASS), PO Box  76, Epping, NSW 1710, Australia}
\author{S.~Murray}
\affiliation{International Centre for Radio Astronomy Research (ICRAR), Curtin University, Bentley, WA 6102, Australia}
\affiliation{ARC  Centre  of  Excellence  for  All  Sky  Astrophysics  in  3  Dimensions  (ASTRO  3D)}
\affiliation{School of Earth and Space Exploration, Arizona  State University, Tempe, AZ 85287, USA}
\author{D.~Pallot}
\affiliation{International Centre for Radio Astronomy Research (ICRAR), University of Western Australia, Crawley, WA 6009, Australia}
\author{B.~Pindor}
\affiliation{School of Physics, The University of Melbourne, Parkville, VIC 3010, Australia}
\affiliation{ARC  Centre  of  Excellence  for  All  Sky  Astrophysics  in  3  Dimensions  (ASTRO  3D)}
\author{M.~Rahimi}
\affiliation{School of Physics, The University of Melbourne, Parkville, VIC 3010, Australia}
\author{J.~Riding}
\affiliation{School of Physics, The University of Melbourne, Parkville, VIC 3010, Australia}
\author{G.~Sleap}
\affiliation{International Centre for Radio Astronomy Research (ICRAR), Curtin University, Bentley, WA 6102, Australia}
\author{K.~Steele}
\affiliation{International Centre for Radio Astronomy Research (ICRAR), Curtin University, Bentley, WA 6102, Australia}
\author{K.~Takahashi}
\affiliation{Kumamoto University, Japan}
\affiliation{International Research Organization for Advanced Science and Technology, Kumamoto University, Kumamoto, Japan}
\author{S.~J.~Tingay}
\affiliation{International Centre for Radio Astronomy Research (ICRAR), Curtin University, Bentley, WA 6102, Australia}
\author{M.~Walker}
\affiliation{International Centre for Radio Astronomy Research (ICRAR), Curtin University, Bentley, WA 6102, Australia}
\author{R.~B.~Wayth}
\affiliation{International Centre for Radio Astronomy Research (ICRAR), Curtin University, Bentley, WA 6102, Australia}
\affiliation{ARC  Centre  of  Excellence  for  All  Sky  Astrophysics  in  3  Dimensions  (ASTRO  3D)}
\author{R.~L.~Webster}
\affiliation{School of Physics, The University of Melbourne, Parkville, VIC 3010, Australia}
\affiliation{ARC  Centre  of  Excellence  for  All  Sky  Astrophysics  in  3  Dimensions  (ASTRO  3D)}
\author{A.~Williams}
\affiliation{International Centre for Radio Astronomy Research (ICRAR), Curtin University, Bentley, WA 6102, Australia}
\author{C.~Wu}
\affiliation{International Centre for Radio Astronomy Research (ICRAR), University of Western Australia, Crawley, WA 6009, Australia}
\author{J.~S.~B.~Wyithe}
\affiliation{School of Physics, The University of Melbourne, Parkville, VIC 3010, Australia}
\affiliation{ARC  Centre  of  Excellence  for  All  Sky  Astrophysics  in  3  Dimensions  (ASTRO  3D)}
\author{S.~Yoshiura}
\affiliation{Kumamoto University, Japan}
\author{Q.~Zheng}
\affiliation{Shanghai Astronomical Observatory, China}






\begin{abstract}
The compact configuration of Phase II of the Murchison Widefield Array (MWA) consists of both a redundant subarray and pseudo-random baselines, offering unique opportunities to perform sky-model and redundant interferometric calibration. The highly redundant hexagonal cores give improved power spectrum sensitivity. In this paper, we present the analysis of nearly 40 hours of data targeting one of the MWA's EoR fields observed in 2016.  We use both improved analysis techniques presented in \cite{barry2019newlimit} as well as several additional techniques developed for this work, including data quality control methods and interferometric calibration approaches. We show the EoR power spectrum limits at redshift 6.5, 6.8 and 7.1 based on our deep analysis on this 40-hour data set. These limits span a range in $k$ space of $0.18$ $h$ $\mathrm{Mpc^{-1}}$ $<k<1.6$ $h$ $\mathrm{Mpc^{-1}}$, with a lowest measurement of $\Delta^2\leqslant2.39\times 10^3$ $\mathrm{mK}^2$ at $k=0.59$ $h$ $\mathrm{Mpc^{-1}}$ and $z=6.5$. 
\end{abstract}


\section{Introduction} 
\label{sec:intro}
The exploration of the Epoch of Reionization (EoR) has the potential to reveal substantial information about the evolution of early universe, as well as the UV and X-ray properties of the first galaxies. One of the most promising tools for direct EoR detection is the 21 cm emission from the hyper-fine level transition in neutral hydrogen. In principle, by observing the redshifted 21 cm signal we are able to image the 3D map of neutral IGM evolution over the history of the universe (for reviews of ``21 cm cosmology," see, e.g., \citealt{furlanetto2006cosmology,morales2010review,pritchard201221}.

However, recovering a complete picture of the EoR by observing the 21 cm signal is rather ambitious due to several challenges inherent to the technique. The greatest obstacle is the overwhelmingly bright galactic and extragalactic foregrounds, which are 4 to 5 orders of magnitude stronger than the target signal \citep{oh2003foregrounds,santos2005foregrounds}. Extracting the cosmological signal in the presence of bright foregrounds will require instruments with extremely high sensitivity, as well as precise data analysis methods \citep{pober2014next,morales2018understanding}.

\begin{figure}
    \centering
    \includegraphics[width=\linewidth]{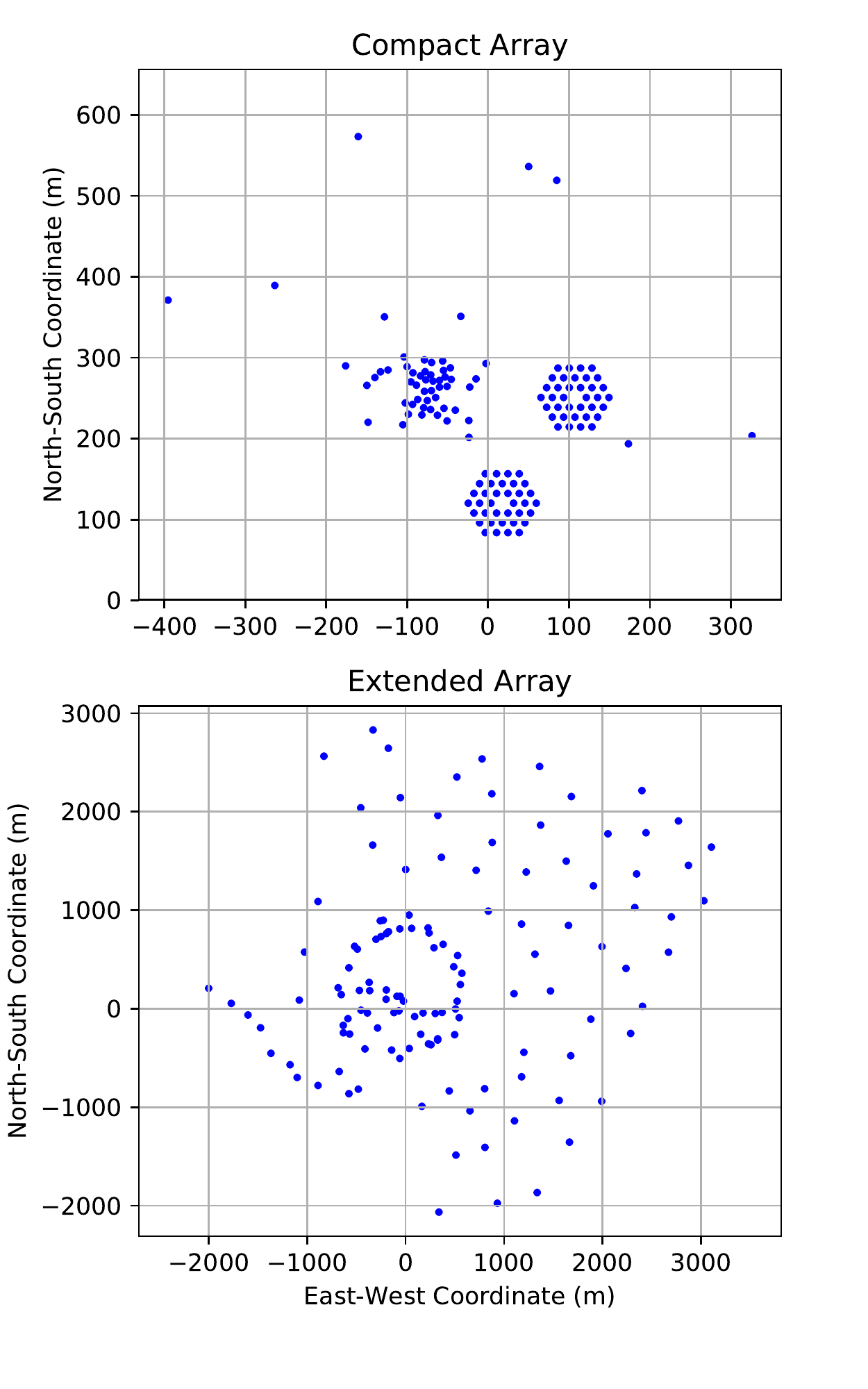}
    \caption{MWA Phase II array layout. Top: Compact Array; Bottom: Extended Array}
    \label{fig:layout}
\end{figure}

In recent years several ground-based experiments with the aim of detecting the 21 cm cosmological signal detection have been underway, such as the Giant Metrewave Radio Telescope (GMRT, \citealt{paciga2013simulation}), the Low Frequency Array (LOFAR, \citealt{van2013lofar}), the Donald C. Backer Precision Array for Probing the Epoch of Reionization (PAPER, \citealt{parsons2010precision}), and the Murchison Widefield Array (MWA, \citealt{tingay2013murchison, bowman2013science,wayth2018phase}). The next generation of telescopes with improved sensitivities are also under construction, such as the Hydrogen Epoch of Reionization Array (HERA, \citealt{pober2014next,deboer2017hydrogen}) and the Square Kilometre Array (SKA, \citealt{mellema2013reionization}). Based on observations that have been conducted by these telescopes, increasingly stringent EoR power spectrum limits have been published \citep{paciga2013simulation,dillon2014overcoming,dillon2015empirical,beardsley2016first,paul2016delay,patil2017upper,barry2019newlimit,kolopanis2019newlimit}; a potential first detection of the sky-averaged 21\,cm signal has also been reported by the Experiment to Detect the Global EoR Signature (EDGES; \citealt{bowman2018edges}.
In 2016 the MWA began Phase II of its experiment, deploying an additional 128 tiles to upgrade the telescope into a 256-element array \citep{wayth2018phase}. Currently, only 128 tiles can be correlated simultaneously, so Phase II operations split time between a compact array and an extended array, each consisting of 128 tiles (Figure \ref{fig:layout}). The compact array has a new array layout with redundant baselines added, which brings in improved power spectrum sensitivity and the capability of redundant calibration. 

In this paper, we present new EoR power spectrum limits at redshifts 7.1, 6.8 and 6.5 based on 2016 observations from the MWA Phase II compact array. We implement the FHD/$\varepsilon$ppsilon interferometric processing pipeline \citep{jacobs2016murchison,barry2019fhd} to analyze our dataset (section \ref{sec:pipe}).
Concurrent with this work, \cite{barry2019fhd} and \cite{barry2019newlimit} developed a number of improvements to the FHD/$\varepsilon$ppsilon pipeline to improve the analysis over that presented in \citealt{beardsley2016first}.  \cite{barry2019newlimit} presents an application of this improved analysis to the same Phase I data set used in \cite{beardsley2016first} to produce a significantly improved limit on the EoR signal strength. In this work, we use several of these new techniques, including the modified uv-space gridding kernel which we review in section \ref{sec:pipe-details}.  We also present several new and alternative techniques developed for this analysis, including data quality metrics and calibration techniques, detailed in sections \ref{sec:qltm} and \ref{sec:cal}, respectively.
Similar to \cite{beardsley2016first} and \cite{barry2019newlimit}, we also compare the results of the FHD/$\varepsilon$ppsilon pipeline with the independent RTS/CHIPS pipeline \citep{mitchell2008real,ord2010interferometric,jacobs2016murchison,trott2016chips} on a subset of our data as a robustness check.

The overall structure of this paper is as follows.  In section \ref{sec:data}, we describe the data used in this analysis.  In section \ref{sec:pipe}, we provide an overview of the FHD/$\varepsilon$ppsilon pipeline, including the developments of \cite{barry2019fhd} and \cite{barry2019newlimit}.
Our first modification to the pipeline is the application of suite of data quality metrics using noise spectrum evaluation \citep{wilensky2019ins}, redundant calibration $\chi^2$ values \citep{zheng2014miteor,li2018comparing}, and EoR window power measurements (similar to \citealt{beardsley2016first}); this is described in section \ref{sec:qltm}. Our second modification is the implementation of a new interferometric calibration approach by combining redundant calibration and sky-model-based calibration \citep{li2018comparing} also using an auto-correlation based bandpass smoothing technique distinct from that presented in \cite{barry2019fhd}; this is described in section \ref{sec:cal}. In section \ref{sec:psalyz}, we detail the specifics of our power spectrum analysis, and we present the first season MWA Phase II power spectrum limits in section \ref{sec:res}.  In section \ref{sec:ps-comparison}, we compare these results with previous MWA Phase I deep integration analyses (\citealt{beardsley2016first} and \citealt{barry2019newlimit}) and investigate the main developments leading to our power spectrum limits.  We conclude in section \ref{sec:conclusion}.

\section{Data Overview}
\label{sec:data}
The Murchison Widefield Array (MWA) is one of the first generation telescopes with the aim of detecting the 21 cm EoR signal. Phase I of the project consisted of 128 pseudo-randomly distributed tiles, covering an area with a diameter of 3 km \citep{tingay2013murchison}. Each MWA tile consists of 16 dual-polarization dipoles arranged in 4$\times$4 layout. By adding analog delays to each dipole, the tile beam can be adjusted to point to a specific target sky field. The MWA EoR project primarily uses a `drift-and-shift' method for observations of a specific field, i.e., the beam pointing direction gets steered in discrete steps approximately every 30 minutes and then observes in a drift-scan mode until the next pointing shift.  We refer to each discrete beam setting as a ``pointing". There are 8 pointings in our data set. We label them in local sidereal time order as -5, -4, -3, -2, -1, 0, 1, 2, where 0 represents the zenith pointing.  

The MWA operates at 80-300 MHz, with a processing bandwidth of 30.72 MHz selected from this larger range. The signals get channelized into 24 coarse channels (or bands) after travelling to the receivers, which introduces 1.28MHz aliasing, thus coarse band edges must be flagged in our analysis \citep{prabu2015digital}. Each coarse band is further channelized into 32 fine frequency channels by the correlator, with a resolution of 40 kHz. 

The MWA Phase II consists of a compact array and an extended array (Figure \ref{fig:layout}), each consisting of 128 tiles; only one configuration can be operated a time, and a non-trivial configuration process is needed to switch between them. The data set we analyze comes from MWA Phase II compact array (top panel of Figure \ref{fig:layout}), which consists of the two hexagonal cores and 56 pseudo-randomly distributed tiles.

In this work, we analyzed 40 hours of EoR0 (RA 0\,h, Dec -27$^\circ$) high band (167 - 197 MHz, corresponding to z = 7.5-6.2 in the 21 cm line) data. In this data set, the observations began on October 15, 2016, and the latest observations were taken on December 15, 2016. Data are observed with a time resolution of 0.5 seconds and a fine frequency resolution of 40 kHz.  All data are divided into observation snapshots, each consisting of 112 seconds of data. The total number of observation snapshots used in our analysis is 1255, corresponding to 40 hours of data.

\section{Analysis Pipeline}
\label{sec:pipe}
In this work, we use the FHD/$\varepsilon$ppsilon analysis pipeline \citep{barry2019fhd}, with modifications to both the data quality metrics and calibration strategy. The software Fast Holographic Deconvolution\footnote{\url{https://github.com/EoRImaging/FHD}} (FHD, \citealt{barry2019fhd}) is developed for interferometric data modeling, calibration and imaging. The package Error Propagated Power Spectrum with InterLeaved Observed Noise\footnote{\url{https://github.com/EoRImaging/eppsilon}} ($\varepsilon$ppsilon, \citealt{jacobs2016murchison,barry2019fhd}) calculates the 21 cm power spectrum based on the image outputs from FHD, with errors propagated through the analysis.

\subsection{Pre-processing}
\label{subsec:prepro}

The preprocessing is done by the \texttt{COTTER} pipeline. \texttt{COTTER} uses \texttt{AOFLAGGER}\footnote{\url{http://aoflagger.sourceforge.net/}} for Radio Frequency Interference (RFI) flagging. The methodology is described in detail in \cite{offringa2010post,offringa2012morphological}. In addition, both sides (80 kHz width at each side) and the center (40 kHz width) of every coarse band are flagged due to aliasing of the poly-phase filter bank and DC offsets, respectively \citep{offringa2015low}. 

\texttt{COTTER} also does data format conversion and data volume reduction. 
The raw data coming out from the correlator is in a non-standard GPU box format. The every 1.28 MHz coarse band data is written into a separate data file. \texttt{COTTER} converts these data files into a more standard readable \texttt{uvfits} format. To reduce the data volume, in this work, the data is averaged from its native 0.5 second time resolution to 2 seconds resolution; the frequency resolution is kept at 40 kHz.

\subsection{FHD/$\varepsilon$ppsilon Pipeline}
\label{sec:pipe-details}

The power spectrum pipeline used in this work principally uses the FHD and $\varepsilon$ppsilon packages. For a fuller description of this pipeline, see \cite{jacobs2016murchison} and \cite{barry2019fhd}; here we present the key details for our analysis.

FHD uses an instrument and sky model to simulate visibilities that are used for sky-model-based calibration (see section \ref{sec:cal}). 
After calibration, we subtract the model visibilities produced by FHD from the calibrated data. We then use FHD to grid the residual visibilities into $(u,v,f)$ space on a per-observation basis.

Note that before doing any gridding, FHD separates each observation snapshot into even and odd time steps and grids these two halves of the data into separate cubes. As described in \cite{barry2019fhd}, we use the even-odd noise calculation to simultaneously check our noise propagation and calculate the cross-power.

In earlier analyses with FHD/$\varepsilon$ppsilon, a kernel corresponding to the Fourier transform of the instrument primary beam was used to grid the visibilities, as this was shown to maximize sensitivity in interferometry analyses \citep{morales2009software}. \cite{barry2019newlimit} demonstrated that applying a squared Blackman-Harris window to the beam image and then generating the uv-kernel (which is effectively a convolution in uv-space) serves to mitigate analysis systematics, including the beam smoothness and gridding resolution issues described in \cite{kerrigan2018improved}.  We also apply this modified gridding kernel in our analysis.

Following the gridding step, FHD Inverse Fourier Transforms the per-observation $(u,v,f)$ cubes to image space. The image cubes from each observation are then projected into a series of HEALPix maps, one for each frequency \citep{gorski1999healpix}. In this analysis, as in \cite{barry2019newlimit}, we generate HEALPix maps that are approximately 10 times larger than \cite{beardsley2016first} to avoid aliasing effects due to a limited extent in the image.

The per-observation interleaved HEALPix cubes\footnote{Strictly speaking, these are not cubes, since the HEALPix coordinate system corresponds to a curved sky. However, we use this nomenclature to refer to data products with two angular coordinates and one frequency axis, even if they are not perfectly rectilinear.} produced by FHD are the input data products for $\varepsilon$ppsilon.
All HEALPix maps from all observations are co-added; by combining observations in the HEALPix frame, we naturally account for wide-field sky curvature effects. The summed image cube (in $(\theta_x,\theta_y,f)$) is then Direct Fourier transformed (DFT) into $(k_x,k_y,f)$ space. 

Next $\varepsilon$ppsilon weights the data using a sampling map generated by FHD. This upweights well-measured modes, downweights poorly measured modes and constructs the variance weighted sum and difference of the even and odd cubes to generate mean and noise cubes \citep[eq. 22]{barry2019fhd}.


As described in \citet{barry2019fhd}, $\varepsilon$ppsilon uses the Lomb-Scargle periodogram  \citep{lomb_least-squares_1976,scargle_studies_1982} to go to $(k_x,k_y,k_z)$ space instead of a direct Fourier transform because of the non-uniform sampling along the frequency direction (caused by the coarse band pass, RFI flagging, and the evolution of baseline lengths with frequency). $\varepsilon$ppsilon also applies a Blackman-Harris window along the frequency axis before calculating the periodogram to achieve a larger dynamic range at the cost of reducing the effective bandwidth by a factor of 2.

Finally, the power spectra are calculated as the mean cube squared minus the difference cube squared \citep[eq. 25]{barry2019fhd}, which is equivalent to the cross-power between the even and odd cubes so it has no noise bias.

\subsection{Unbiased Estimator}
\label{sec:unbiased}

We wish to confirm that the FHD/$\varepsilon$ppsilon pipeline indeed provides an unbiased estimate for the power spectrum (e.g. there is no ``signal loss").  \cite{barry2019newlimit} conduct an end-to-end pipeline simulation and confirm that, for the Phase I baseline layout, FHD/$\varepsilon$ppsilon recovers the correct power spectrum for an input EoR signal.  We reproduce this result exactly with our analysis pipeline.

However, \cite{barry2019fhd} demonstrates that the overall power spectrum normalization is affected by the density of baseline sampling in the uv plane.  The normalization needs to account for the covariance between visibilities that overlap during the gridding operation \citep{liu2014epoch}.  Carrying around the full visibility covariance matrix is not computationally tractable for FHD/$\varepsilon$ppsilon, and so this effect is handled numerically.  Simulations show that the correction to the normalization asymptotes to a factor of 2 for dense baseline sampling.  This effect can be thought of as an increase of the effective area in the uv plane contributing any uv location; see Appendix A of \cite{barry2019fhd} for more details.

This factor of 2 normalization scaling is included in the calculation of power spectra from MWA Phase I, given its smooth and dense baseline sampling.  However, while the redundant layout of the Phase II array still produces a filled uv plane, the sampling is far from uniform---most of the sensitivity comes from the handful of points in uv plane sampled by the large number of redundant baselines.  However, earth rotation smooths out uv coverage of the Phase II array, bringing it into the same sampling limit as Phase I, which we confirm through simulation.  We therefore apply the same factor of 2 normalization scaling as described in \cite{barry2019fhd} to our final limit analysis in order to return an unbiased power spectrum estimate.  Shorter analyses with Phase II (e.g. a zenith-pointing subset) will, however, require precision simulations to determine the correct normalization.

\subsection{Pipeline Structure}

\begin{figure}
    \centering
    \includegraphics[width=0.8\linewidth]{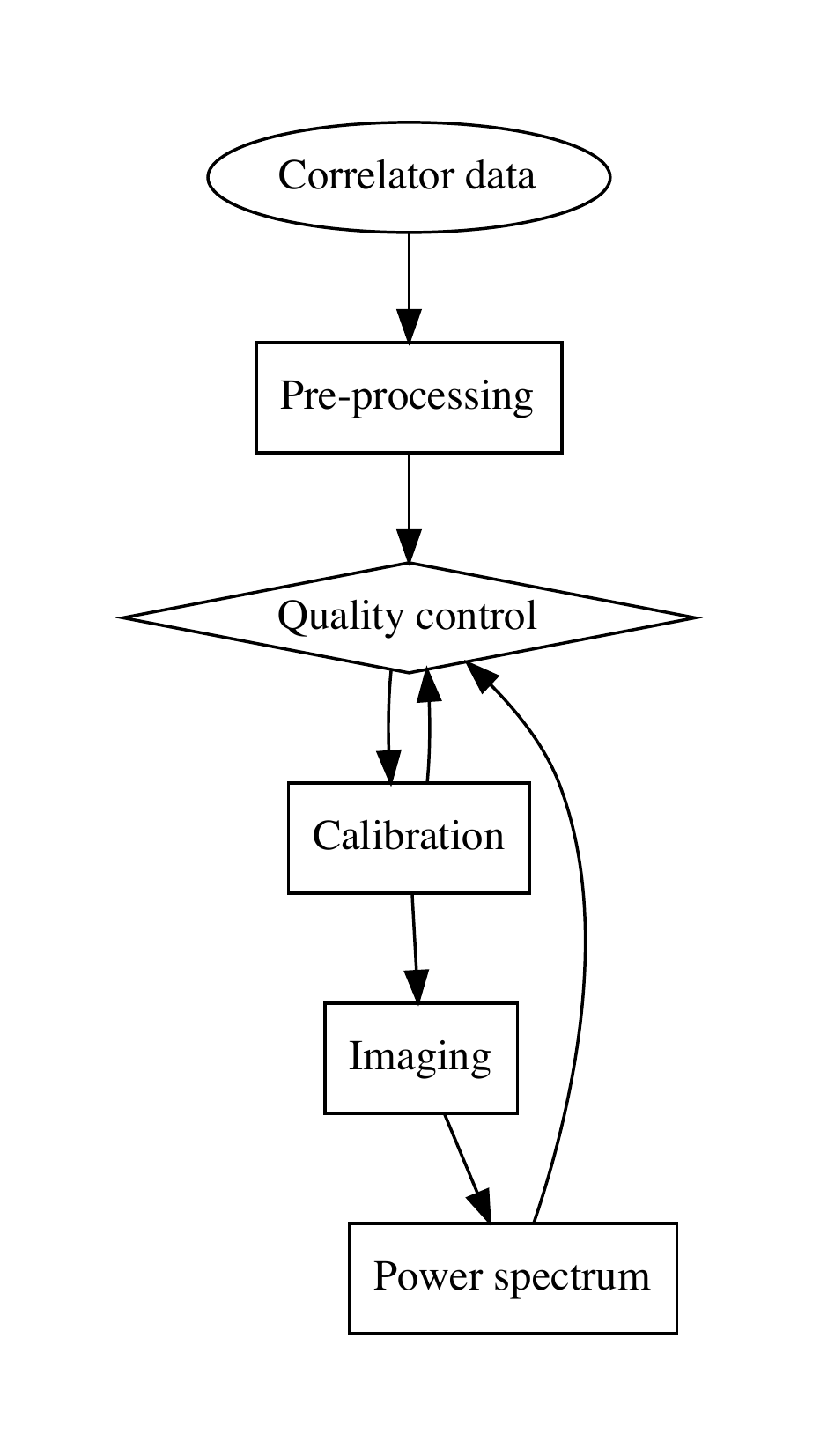}
    \caption{Schematic illustration of the analysis pipeline used in this work.}
    \label{fig:pipe}
\end{figure}

The overall analysis pipeline used in our work is schematically illustrated in Figure \ref{fig:pipe}. Instead of processing data through the pipeline one time, we also evaluate data products from subsequent steps, including calibration solutions and power spectra, and execute further flagging based on these evaluations. Section \ref{sec:qltm} will describe quality metrics that we use for data flagging. 
The other major changes from the pipeline described in \cite{barry2019fhd} and \cite{barry2019newlimit} are contained in the calibration stage.

\section{Data Quality Metrics}
\label{sec:qltm}

As mentioned in section \ref{subsec:prepro}, RFI flagging is performed in \texttt{COTTER} pre-processing pipeline. However, close inspection of the data reveal faint RFI not captured by \texttt{AOFLAGGER}; in this section, we introduce 3 additional metrics used to flag contaminated data. 

\begin{figure}
    \centering
    \includegraphics[width=\linewidth]{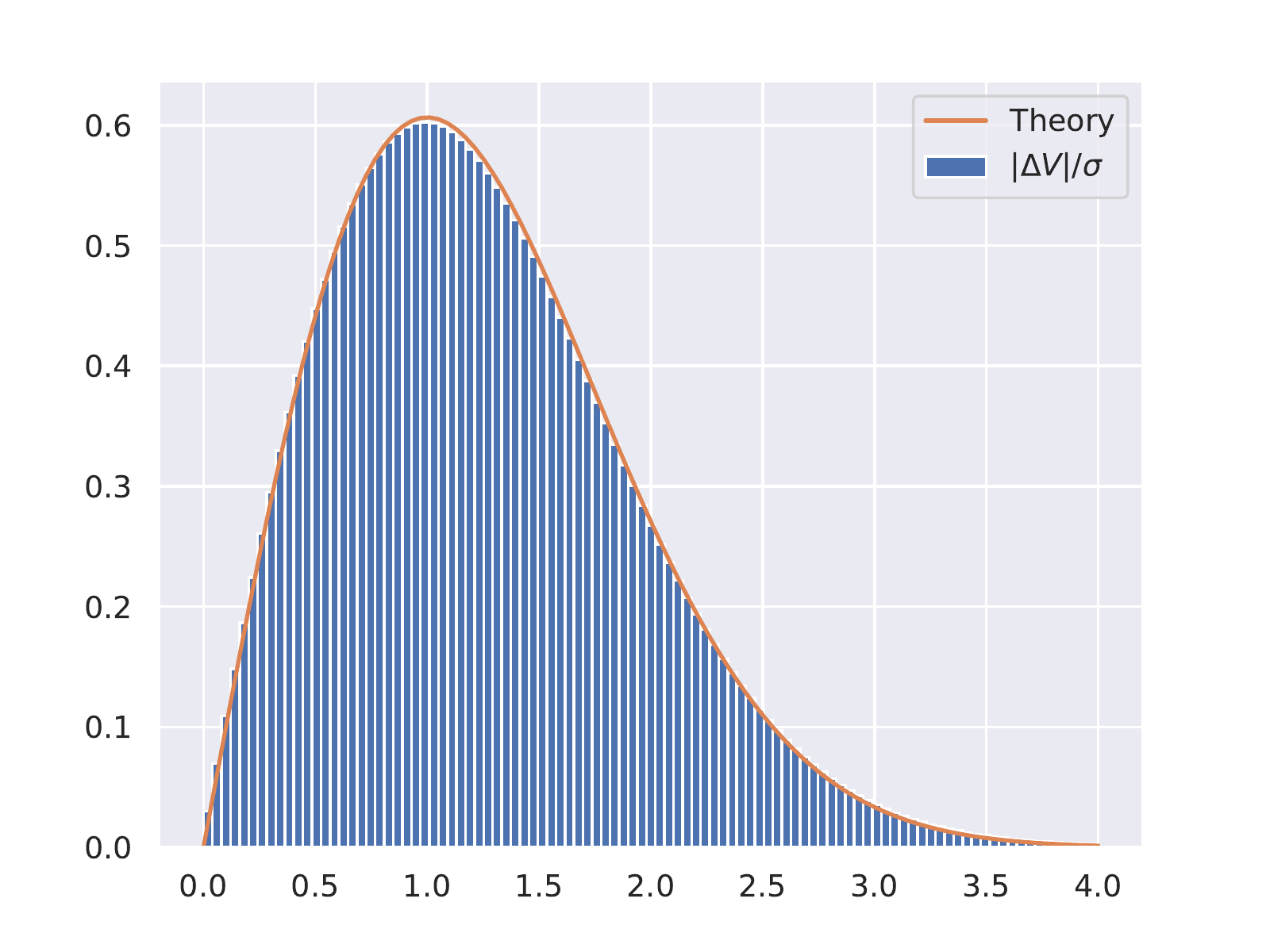}
    \caption{The normalized histogram of $|{\Delta}V|/\sigma$ calculated from one 112 second observation snapshot, where $\Delta V$ is defined in equation \ref{eq:INS}, and $\sigma$ is the noise standard deviation. The orange line shows the theoretical probability density function. This agreement shows that without RFI contamination, the sky-subtracted noise spectrum is Gaussian random distributed with 0 mean.}
    \label{fig:deltav}
\end{figure}

\begin{figure*}
    \centering
    \includegraphics[width=\linewidth]{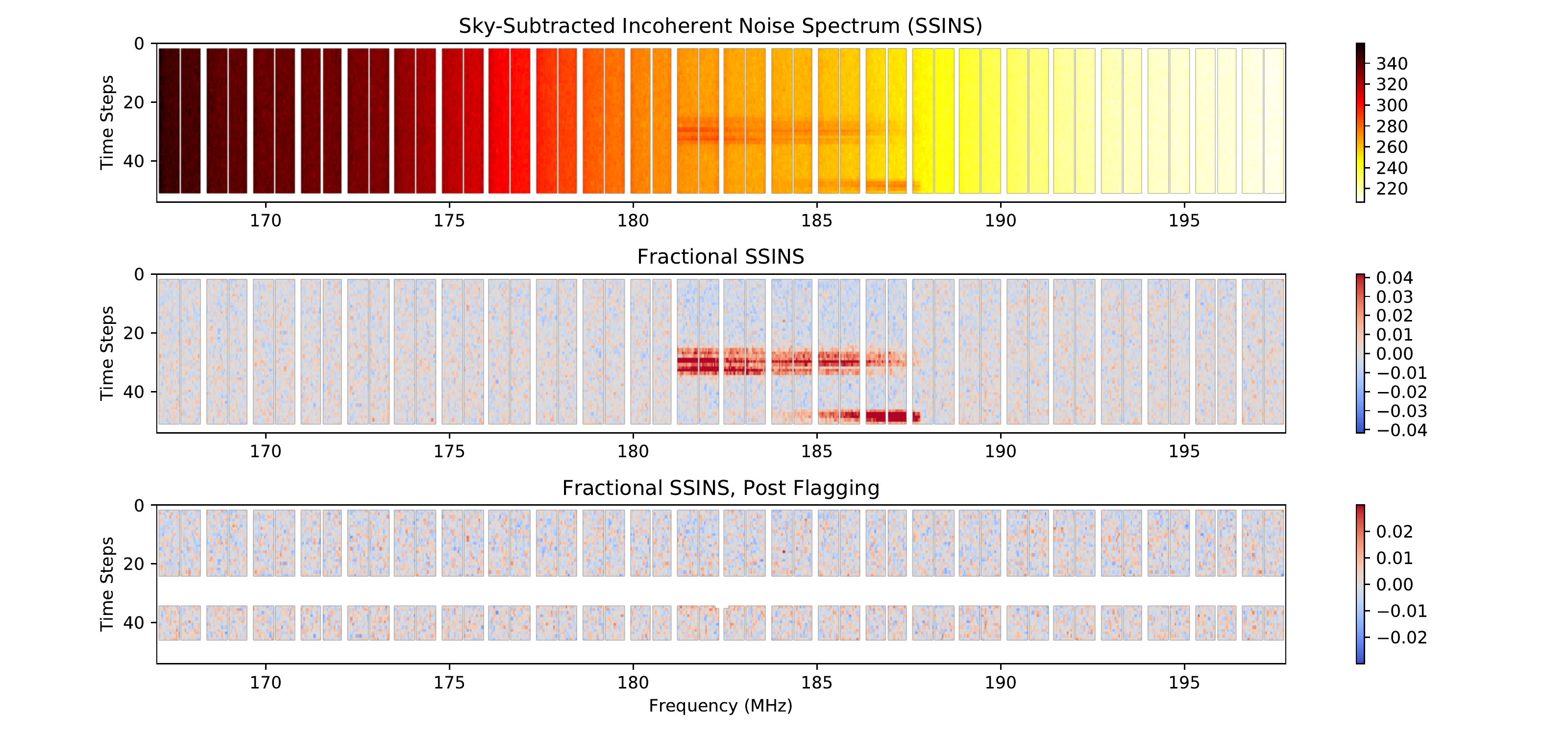}
    \caption{RFI flagging based on SSINS. White pixels are flagged samples. Top: the SSINS calculated from a 112 second observation snapshot in east-west polarization; Middle: the fractional SSINS; Bottom: the fractional SSINS after flagging using the filtering method developed in this work (section \ref{subsec: ins}).}
    \label{fig:ins}
\end{figure*}

\subsection{Sky-Subtracted Incoherent Noise Spectrum}

\label{subsec: ins}

\cite{wilensky2019ins} introduces a metric which can identify faint RFI below the noise level of an individual baseline---the Sky-Subtracted Incoherent Noise Spectrum (SSINS)---which we review here.  This technique begins by differencing visibilities from adjacent time pairs:
\begin{equation}
    \Delta V_{ij}(f,t)=V_{ij}(f,t+\Delta t)-V_{ij}(f,t),
    \label{eq:INS}
\end{equation}
where $V_{ij}(f,t)$ denotes the visibility measured by baseline $ij$ at time $t$ and frequency $f$, and $\Delta t$ is time resolution. As the sky only rotates 15 arcseconds within $\Delta t=2$ seconds, the remnant of sky signal in this subtraction is negligible compared to noise. This step is therefore the ``sky subtraction" aspect of SSINS. If ${\Delta}V$ is purely noise, which is assumed to be Gaussian random distributed complex numbers with 0 mean, $|{\Delta}V|/\sigma$ will follow a Rayleigh distribution, where $\sigma$ is the noise standard deviation. 
An example of histogram of $|{\Delta}V|/\sigma$ of one 112 second observation snapshot at east-west polarization is shown in Figure \ref{fig:deltav}. The ``incoherent" aspect of SSINS comes from averaging the amplitudes of time pair differences over all baselines as a function of time and frequency:
\begin{equation}
    \langle|{\Delta}V(f,t)|\rangle=\frac{1}{N_\mathrm{bls}}\sum_{ij}{|{\Delta}V_{ij}(f,t)|},
    \label{eq:INSmean}
\end{equation}
where $N_\mathrm{bls}$ is the number of baselines. Equation \ref{eq:INSmean} is the SSINS. According to the central limit theorem, $\langle|{\Delta}V(f,t)|\rangle$ follows a normal distribution as the number of baselines is large ($N_\mathrm{bls} \approx$ 8000). An example of SSINS from an observation snapshot is shown in the top panel of Figure \ref{fig:ins}. The vertical white streaks are flagged frequency channels due to coarse band systematics as we mentioned in section \ref{subsec:prepro}.

\cite{wilensky2019ins} also describes a method for flagging RFI contaminated data based on SSINS.  We concurrently developed an alternative approach for flagging based on SSINS, which we describe here.  We refer the reader to \cite{wilensky2019ins} for a description of their method.  Ultimately, we expect the differences between the two flagging approaches to be small, but we leave a detailed comparison for a future work.

To remove the frequency dependent noise level in the SSINS (due to the spectral slope of the synchrotron dominated sky noise and instrument band structure), we divide out the median value of each frequency channel in the SSINS. We then subtract 1 from it, leading to the fractional SSINS, as shown in the middle panel of Figure \ref{fig:ins}. After this scaling and shifting, if there is no RFI, all time/frequency bins of the fractional SSINS are independent and identically distributed Gaussian random variables. 

In the example of Figure \ref{fig:ins}, the data has already passed the \texttt{AOFLAGGER} step. However, there is still some faint non-Gaussian structure in the center of the band. The feature in the center of the band (181 - 188 MHz) is due to a typical TV channel contamination. To identify this and similar RFI, we first recursively flag the time/frequency bins which are 5 standard deviation outliers from the mean. However, the feature is not completely flagged simply by finding outliers. We identify fainter components of the feature using the coherence of time/frequency bins along the frequency axis. We convolve the SSINS shown in the middle panel of Figure \ref{fig:ins} with a Gaussian kernel with a FWHM of 640kHz, which down weights noise like areas and up weights RFI features. We then flag regions in the convolved SSINS using 5 standard deviation threshold. We also check the flagging levels within each coarse band for each time pair; if over half the fine frequencies in a coarse channel are flagged in a given time pair, we flag the whole coarse band. Finally, we aggressively flag entire time pairs which have more than 2 coarse bands flagged in previous steps. The bottom panel of Figure \ref{fig:ins} shows the resultant flagging after all of these steps. For the final analysis, we completely remove any observation snapshots where more than 50\% of the data is flagged in this procedure. Overall, this metric flags roughly an extra 5.6 hours of data in addition to \texttt{AOFLAGGER}. 
\begin{figure*}
    \centering
    \includegraphics[width=\linewidth]{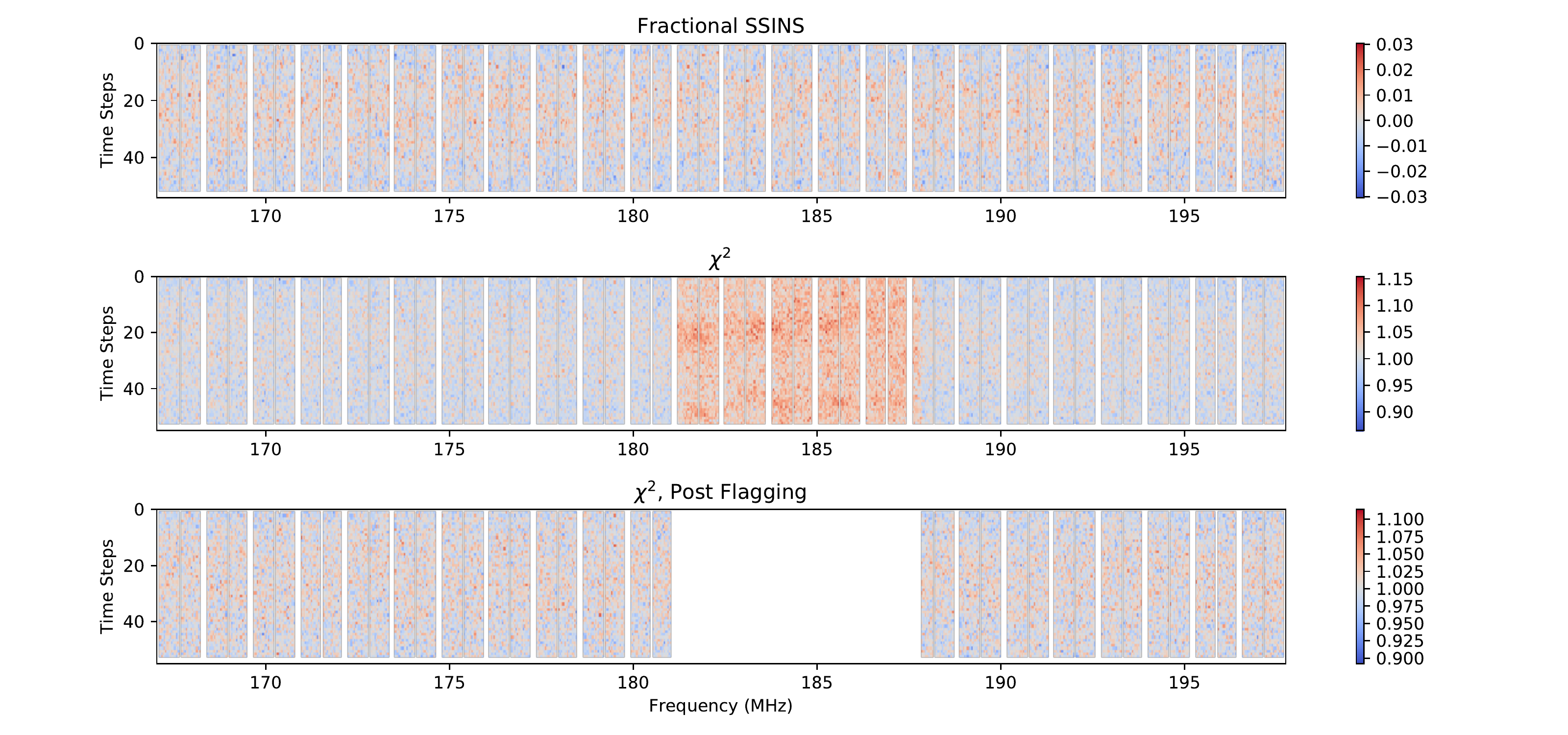}
    \caption{Faint RFI detection using redundant calibration $\chi^2$ (middle panel), comparing with SSINS from the same data set (top panel). The bottom panel shows the $\chi^2$ with RFI flagged. The $\chi^2$ metric is able to detect RFI that is missed by the SSINS.}
    \label{fig:chi-ins}
\end{figure*}

\subsection{$\chi^2$ of Redundant Calibration}
\label{subsec: redchi}
There also exists RFI that is detected by only part of the array\footnote{We found this is due to the transmitter being near the horizon in the south as we imaged the RFI contaminated data.}; such emission introduces non-identical signals received by individual tiles. Since SSINS takes an incoherent average over baselines, it can be less sensitive to RFI that is not seen on every baseline. We also find this type of RFI is often time-stable.  The time pair subtraction in SSINS can therefore cancel some of the RFI out. Utilizing the redundant layout of the MWA Phase II compact array, however, this RFI can be detected by $\chi^2$ evaluation in redundant calibration \citep{zheng2014miteor}, which will be described in section \ref{subsec: redcal}. The $\chi^2$ is defined in Equation \ref{eq:chisq}, evaluating the agreement between measurements from redundant baselines. 

In the case where RFI breaks the redundancy in signals received by tiles, visibilities measured by redundant baselines, i.e., baselines with the same length and orientation, will show different behaviors, making the $\chi^2$ value for that baseline type larger. The $\chi^2$ and SSINS of an example observation snapshot is shown in Figure \ref{fig:chi-ins}. This shows a comparison between fractional SSINS (top) and redundant calibration $\chi^2$ (middle) over time and frequency calculated from the same observation (note this is not the same observation shown in Figure \ref{fig:ins}).
The $\chi^2$ indicates biased features between 181 MHz and 188 MHz, which are not detected by SSINS. We flag this channel range and show the $\chi^2$ after flagging in the bottom of Figure \ref{fig:chi-ins}.
We note that this type of RFI also appears highly polarized and is principally detected in the East-West polarization, especially on the nights of October 30, 2016 and November 19, 2016, starting from local sidereal time of 1h. We remove any observation snapshots with anomalous $\chi^2$ features from our final analysis. In this metric, there are 0.8 hours of East-West polarization data being flagged in addition to \texttt{AOFLAGGER} and SSINS. 

\subsection{EoR Window Power}
\label{subsec: wpmetric}
\begin{figure}
    \centering
    \includegraphics[width=0.5\linewidth]{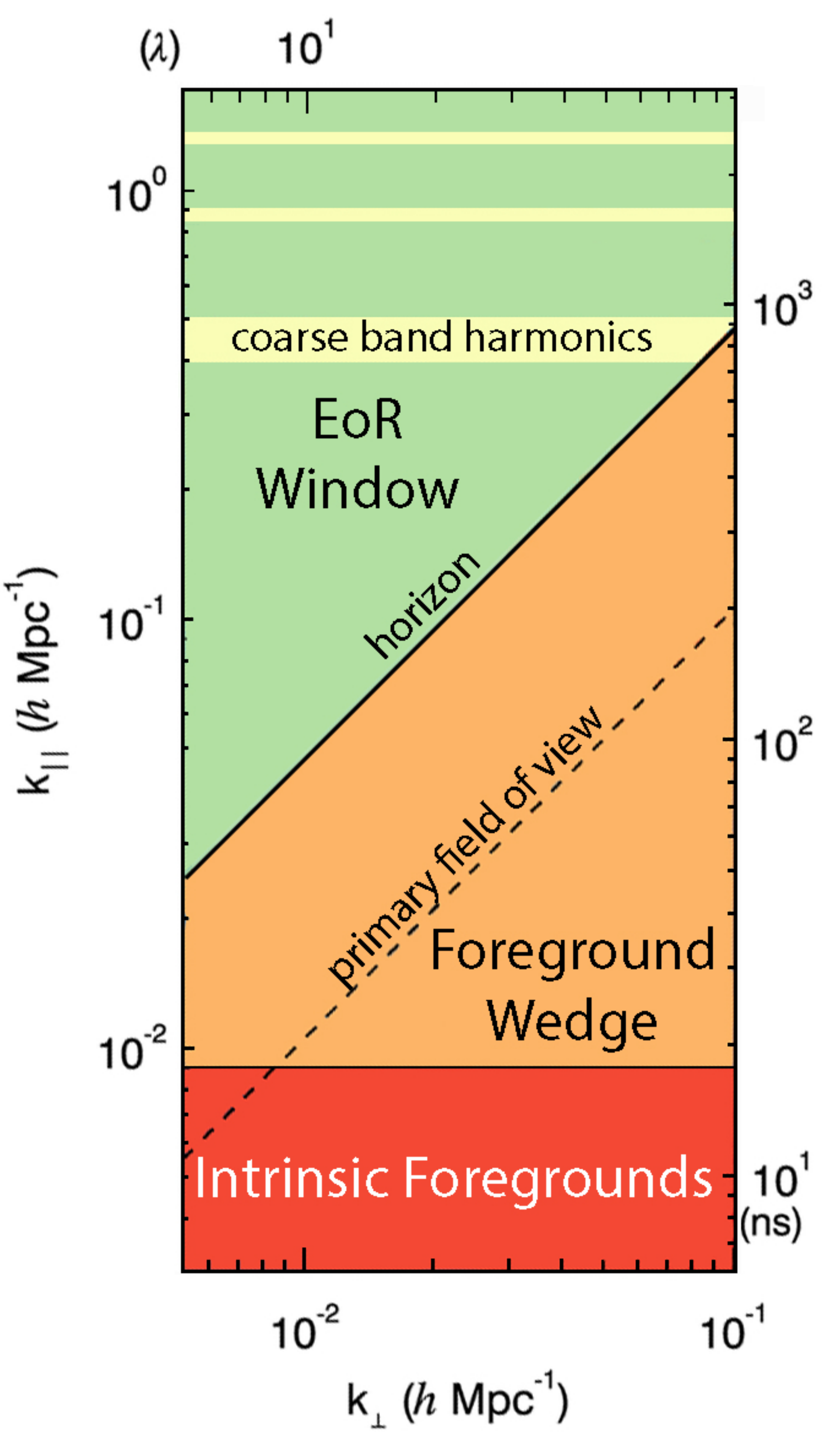}
    \caption{A schematic of an MWA 2-D power spectrum. The lowest $k_\parallel$ modes are dominated by foregrounds (red). The intrinsic chomaticity of interferometry mixes foreground contamination up to high $k_\parallel$ modes which forms the foreground wedge (orange). The horizon line (solid) and the primary field of view line (dashed) mark the upper bound of modes contaminated by foregrounds within the horizon and within the primary field of view, respectively. The remaining `EoR window' (green) is ideally free from foreground contamination. The horizontal streaks (yellow) are coarse band contamination specific to the MWA instrument due to the missing data in coarse band edges.}
    \label{fig:psmodel}
\end{figure}

\begin{figure}
    \centering
    \includegraphics[width=\linewidth]{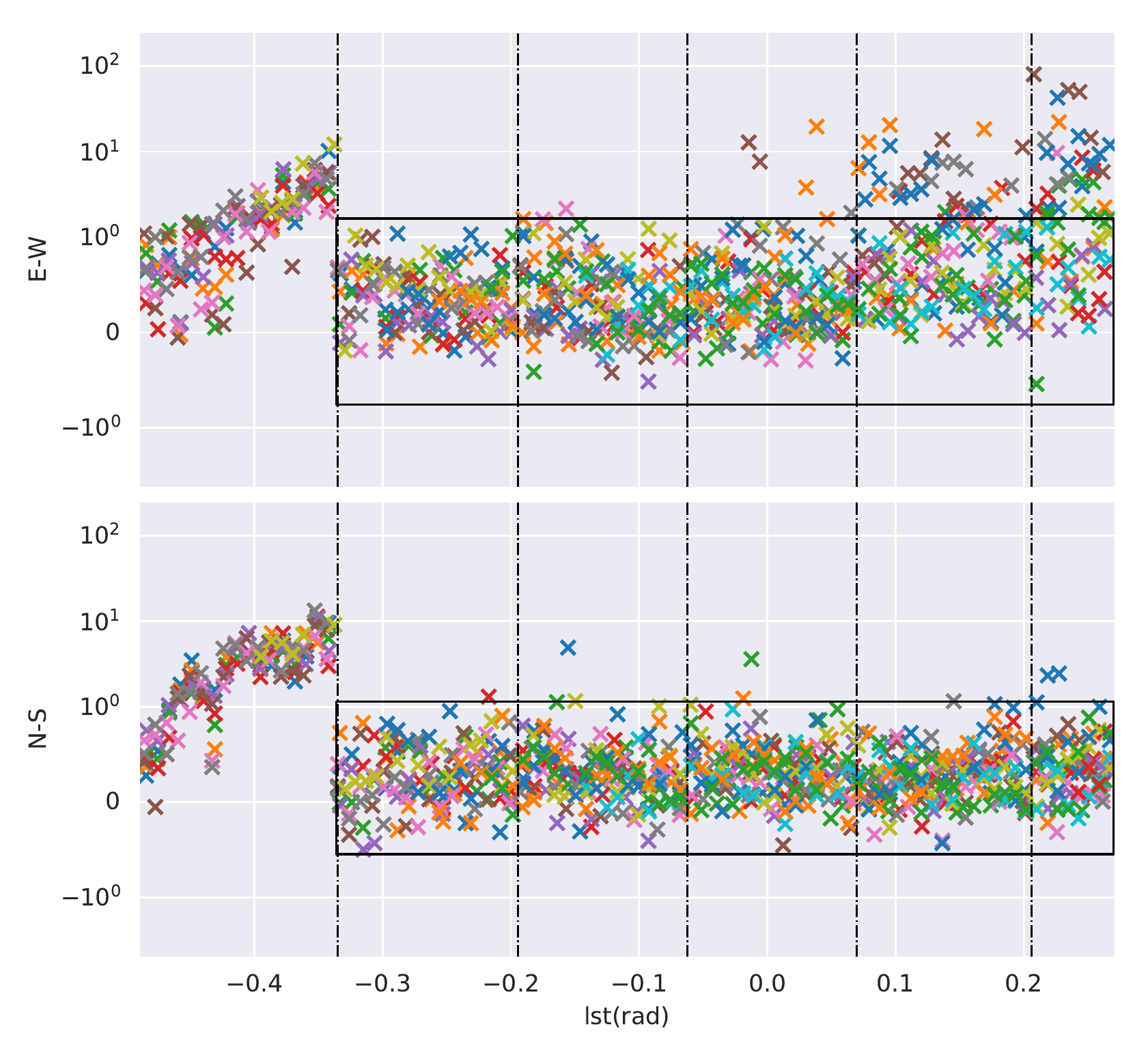}
    \caption{The averaged window power divided by noise level for each individual 2 minute observation at East-West polarization (top) and North-South polarization (bottom) versus local sidereal time (lst) in radians. The vertical dash--dotted lines separate different pointings (-3 to +2, from left to right). Different color symbols represent observations taken from different days. The boxes indicate the observations selected for our final analysis.}
    \label{fig:wp}
\end{figure}

As we mentioned in section \ref{sec:data}, there are 8 pointings in our data set. We find that the calibration (see in section \ref{sec:cal}) does not converge for observations within pointings -5 and -4. In these two pointings, the bright Milky Way galaxy enters the sidelobes of the MWA primary beam, introducing strong foregrounds that are not in included in our sky model, limiting the effectiveness of calibration \citep{beardsley2016first}. We exclude pointings -5 and -4 (7.2 hours of data) for the remainder of this analysis. For pointings -3 through 2, calibration converges and we can apply the `EoR window power' metric we now describe.

Our goal is to obtain a power spectrum as illustrated in Figure \ref{fig:psmodel} (see \citealt{morales2018understanding} for a pedagogical description of the power spectrum features). The foregrounds are predominantly synchrotron and bremsstrahlung emission, which are spectrally smooth, while the cosmological signal is fluctuating rapidly across frequency since each frequency corresponds to distinct  spatial structures at different redshifts. The lowest $k_\parallel$ modes, which are calculated from a Lomb--Scargle periodogram along the frequency axis of the data, are therefore dominated by spectrally smooth foregrounds (red in Figure \ref{fig:psmodel}). The intrinsic frequency dependence (often referred to as chromaticity) of an interferometer mixes foreground modes up to high $k_\parallel$ modes, producing the `foreground wedge' (orange). The remaining `EoR window' (green) is expected to be foreground free and contains signals produced by 21 cm emission from the EoR \citep{datta2010bright,morales2012four,vedantham2012imaging,parsons2012per,trott2012impact,hazelton2013fundamental,thyagarajan2013study,pober2013opening,liu2014epoch}. 
The horizontal streaks are specific to the MWA instrument, which has periodic gaps in its frequency sampling introduced by the flaggings at coarse band edges, as mentioned in section \ref{sec:data}. 

Given our expectation that the EoR window should be free from foregrounds, especially for power spectra made from small amounts of data, we flag individual observations which have significant power contamination in this region. To do this, we run our full analysis on each 112 second observation, producing a 2-D power spectrum where we can evaluate the window power. This is similar to the window power cut in \cite{beardsley2016first}, with a distinction that we run the calibrated data through the actual pipeline to obtain the power spectra instead of calculating delay spectra from raw data. 

After we flag data based on the SSINS and redundant calibration $\chi^2$ metrics, as described in sections \ref{subsec: ins} and \ref{subsec: redchi}, we perform FHD sky-model based calibration on the RFI cleaned data, subtract the foreground model visibilities to create a residual data set, and then run the FHD-$\varepsilon$ppsilon power spectrum pipeline on each residual observation snapshot. 
From the resultant power spectra, we calculate the averaged window power above the horizon line and below the first coarse band harmonic contamination. We divide this value by the noise level propagated to the power spectrum calculated by $\varepsilon$ppsilon for each polarization of each observation. We plot this metric versus local sidereal time for pointing -3 to 2 in Figure \ref{fig:wp}. 
Vertical lines separates different pointings. For pointing -3, we can see this ratio is highly dependent on local sidereal time, suggesting that unmodeled diffuse emission from the sky is contaminating the calibration and residuals, as was the case for pointings -5 and -4. This is similar to the behavior of the MWA Phase I data set presented by \cite{beardsley2016first}. Therefore we remove all observations from pointing -3 from our final analysis as well, which accounts for 4.4 hours of data. Pointings -2, -1, and 0 generally seem consistent over time and polarization, but for pointings 1 and 2, the two polarizations show different behaviors. To identify outlier observations, we evaluated the statistics of the power over noise ratio in Figure \ref{fig:wp} for each polarization over all days, and flag all observations exceeding 3 standard deviation threshold. The black boxes encloses the observations that we select. In addition to per 2 minute observation power spectra, we also evaluate per day per pointing integrated power spectra. We flag observations from a pointing of a day if the integrated power spectrum is a 5 standard deviation outlier comparing with other days. In addition to SSINS and redundant calibration $\chi^2$, this metric flags another 3.2 hours of the East-West polarization data and 0.1 hours of the North-South polarization data. 

After all three quality metrics have been applied, we retain 19 hours of data for the East-West polarization, and 23 hours of data for the North-South polarization.

\section{Calibration}
\label{sec:cal}
Tiles have different gain amplitudes and signal paths, therefore they introduce complex gains to the observed visibilities. The calibration procedure is defined as finding the complex instrument gains from the measured visibilities and removing them from the data. Precise instrument calibration plays a pivotal role in recovering the cosmological signal \citep{barry2016calibration,patil2016systematic,trott2016spectral}. 
Assuming the complex gains ($g_i$'s) are tile based, the relation between the measured visibilities $V_{ij}$ and the true visibilities $y_{ij}$ as a function of time ($t$), frequency ($f$) and polarization ($p$) is described as:
\begin{equation}
\label{eq:cal}
    V_{ij}(t,f,p)\approx g_i(t,f,p)g_j^*(t,f,p)y_{ij}(t,f,p)+n_{ij}(t,f,p),
\end{equation}
where $n_{ij}$ denotes a Gaussian random noise term. For convenience, in the rest of the section, we only discuss the frequency dependence of the gains. The gains are calculated on a per 112 second observation basis, and they are assumed as time-independent in such a short time interval. The calibration is performed independently in East-West and North-South polarization. Both polarizations are calibrated using the same methodology, thus we do not show the polarization variable explicitly in the discussion. 

In this section, we describe our calibration as two major steps: \textbf{per frequency calibration} and \textbf{bandpass calibration}. In the per frequency calibration (described in section \ref{sec:perfreq}), the gains at each frequency are calibrated independently. No correlation between frequency channels is considered. Bandpass calibration (described in section \ref{sec:bpauto}) is performed after the per frequency calibration. In bandpass calibration, we constrain a smooth frequency structure of the gains in order to mitigate contamination due to errors in sky-based and redundant calibration. A concise summary of our final calibration method can be found in section \ref{sec:calsum}.

\subsection{Per Frequency Calibration}
\label{sec:perfreq}
In \cite{li2018comparing}, we combined sky-based calibration and redundant calibration. In this work, we apply a hybrid calibration approach as the per frequency calibration to our data set. Here we briefly review these methodologies. In this subsection, all equations are per frequency based, thus we do not show frequency variable explicitly.

\subsubsection{Sky-based Calibration}
\label{subsec: skycal}
The idea of sky-model-based calibration is to determine complex gains by minimizing the difference between the data and a set of model visibilities. FHD generates model visibilities using a model of the MWA primary beam, the positions of the antennas, and a catalog of the brightness and locations of radio sources in the sky (hereafter referred to as the ``sky model"; \citealt{barry2019fhd}). The sky model comes from the Galactic and Extra-galactic All-sky MWA survey (GLEAM; \citealt{hurley2016galactic}), with approximately 10,000 point sources in the field of view for EoR0 observations. The MWA beam model is generated from the average embedded element model from \cite{sutinjo2015understanding}. The algorithm used by FHD to create model visibilities is known to produce spurious small scale spectral structure at the $\sim$ 1-part-in-$10^3$ level \citep{kerrigan2018improved}; to avoid biases that might result from this effect, we apply a low-pass delay filter to our model visibilities, removing this structure \citep{kerrigan2018improved}.

With model visibilities $m_{ij}$ in hand, we solve for the gains by minimizing the $\chi^2$ defined as:
\begin{equation}
\label{eq:skychi}
    \chi^2=\sum_{ij}\frac{|V_{ij}-g_ig_j^*m_{ij}|^2}{\sigma_{ij}^2},
\end{equation}
where $\sigma_{ij}^2$ denotes the noise variance of baseline $ij$.

To minimize the $\chi^2$ defined in equation \ref{eq:skychi}, FHD uses the StEFCal algorithm described in \cite{salvini2014fast,mitchell2008real}. The resulting products are per-antenna, per-frequency, per-polarization, per-observation complex gains.

\subsubsection{Redundant Calibration}
\label{subsec: redcal}
Redundant calibration requires the tiles to be placed on precise grids so that there are multiple copies of baselines with the same length and orientation \citep{liu2010precision}. In principal, redundant baselines will measure the same Fourier mode of the sky brightness distribution. In a highly redundant array, the number of unknown true visibilities ($y_{ij}$'s) and gains ($g_{i}$'s) are smaller than the number of visibility measurements, which leads to over-determined equation sets. We can then solve for the gains and true visibilities by minimizing the $\chi^2$ defined as:
\begin{equation}
    \chi^2=\frac{1}{N_\mathrm{rbls}-N_\mathrm{urbls}}\sum_{\alpha}\sum_{ij}\frac{|V_{ij}-g_ig_j^*y_\alpha|^2}{\sigma_{ij}^2},
    \label{eq:chisq}
\end{equation}
where $\alpha$ denotes the index of a unique baseline type, $N_\mathrm{rbls}$ is the number of redundant baselines, and $N_\mathrm{urbls}$ is the number of unique redundant baseline types.

The redundant calibration approach consists of two algorithms: Logrithmic calibration (logcal) and Linearized calibration (lincal) \citep{liu2010precision}. Both algorithms are different implementations of a linearization of Equation \ref{eq:cal}. 

In logcal, we take the logarithm of equation \ref{eq:cal} and separate the real and imaginary parts:
\begin{equation}
    \systeme*{\ln(|V_{ij}|)=\ln(|g_i|)+\ln(|g_j|)+\ln(|y_\alpha|)+\Re(\omega_{ij}),\arg(V_{ij}) = \arg(g_i)-\arg(g_j)+\arg(y_\alpha)+\Im(\omega_{ij})},
    \label{eq:logcal}
\end{equation}
where $\omega_{ij}=\ln(1+\frac{n_{ij}}{g_ig_j^*y_\alpha})$ is the noise contribution term.

In lincal, if we have a decent guess for gains (denoted as $g_i^0$) and true visibilities (denoted as $y_\alpha^0$), we can perform Taylor expansion: 
\begin{equation}
    V_{ij}\approx g_i^0g_j^{0*}y_\alpha^0+g_j^{0*}y_\alpha^0\Delta g_i+g_i^0y_\alpha^0\Delta g_j^{*}+g_i^0g_j^{0*}\Delta y_\alpha.
    \label{eq:lincal}
\end{equation}
This again linearizes equation \ref{eq:cal} and we are able to solve for the first order terms $\Delta g_i$'s and $\Delta y_\alpha$'s.

The logcal algorithm is biased in that although $n_{ij}$ is Gaussian distributed with 0 mean, this is not the case in log space ($\omega_{ij}$). lincal is unbiased because the calibration is performed in real-imaginary space. We first perform logcal to obtain a potentially biased but relatively accurate guess for the solutions, then feed them into lincal, solve for the first order parameters, update our solutions, and feed them back to lincal, repeat this process iteratively until the solutions converge. 

We use the package \texttt{omnical}\footnote{\url{https://github.com/jeffzhen/omnical}} \citep{zheng2014miteor} for redundant calibration implementation. 
Redundant calibration also has 4 degeneracy parameters per frequency that it cannot solve for, including 1 absolute amplitude, 1 absolute phase and 2 rephasing degeneracies corresponding to tip and tilt of the array \citep{liu2010precision,zheng2016brute,dillon2018polarized}. We use the degeneracy projection method we introduced in \cite{li2018comparing} to constrain degeneracy parameters. 

\subsubsection{Hybrid Calibration} 
In this work, we combine sky-based calibration and redundant calibration in the following way: we first perform per frequency sky-based calibration using FHD and directly apply the results to the data, then implement \texttt{omnical} on the FHD calibrated data. The latter produces small corrections to the gains for tiles in the hexagonal sub-arrays. In the \texttt{omnical} step, we set degeneracy parameters of redundant calibration to be 0's \citep{li2018comparing}, meaning that the values for these parameters are set only by sky-based calibration.  \cite{byrne2019fundamental} demonstrate that this approach can lead to the same sky-model incompleteness errors described in \cite{barry2016calibration} affecting the power spectrum.  The bandpass calibration described in section \ref{sec:bpauto} can mitigate this error, and future work will investigate ways to further minimize the effect.

\subsection{Bandpass Calibration}
\label{sec:bpauto}

We know that empirically derived calibration solutions will not be perfect. Our sky model and instrument model are not exactly correct, and redundant calibration may introduce errors from imperfectly redundant tile positions and tile beam variations. All these errors may introduce extra spectral structure into the calibration solutions, contaminating the EoR window in the power spectrum \citep{barry2016calibration,joseph2018bias,byrne2019fundamental}.
To mitigate these errors, constraining a smooth bandpass response is a necessity \citep{barry2016calibration}. \cite{beardsley2016first} and \cite{ewall2016first} used a cable-averaged bandpass along with polynomial fitting for the phase and residual gain amplitude to fit the calibration. The cable-averaged bandpass is defined as the averaged gain amplitude among tiles with the same cable length connecting from the tile beamformer to the receiver\footnote{In Phase II compact array, there are 4 cable types: Cables of length 90, 150, and 230 meters
are RG-6, and 320 meters are LMR400-75.}. 
The assumption is that tiles having the same cable type should have the same bandpass structures. Empirically, it is clear that tiles with different cable types do not have the same bandpass structure, at least not at the precision required by an EoR experiment.  Even restricting the bandpass averaging to only antennas with the same cable type, however, has the risk of averaging out gain frequency structures because variations among these tiles are not considered. 
In this work, we propose a new approach to better capture tile to tile variation in the gains and simultaneously improve the smoothness of the bandpass by integrating information of antenna auto-correlations into our bandpass calibration.\footnote{In interferometry, an ``auto-correlation" refers to the signal from one antenna cross multiplied by itself, to distinguish it from a cross-correlation or one antenna's signal cross-multiplied by another's.  Auto-correlations are purely real numbers and represent the total power seen by that antenna as a function of frequency, with no information about the spatial distribution of the incoming signals.}

Using the auto-correlations of MWA tiles to improve calibration was explored by both \cite{ewall2016first} and \cite{barry2019fhd}. Auto-correlations are appealing tools for bandpass calibration for two reasons: they are smooth over frequency due to high SNR and they contain wealth of information about the gain amplitude for each tile. However, it can be challenging to separate the intrinsic foreground spectrum from the instrument bandpass without somehow modeling the foregrounds. Auto-correlations also contain a noise bias, since the system noise is multiplied by itself, leading to a positive-definite offset.  We develop a new auto-correlation calibration technique designed to avoid these issues: instead of fitting the foregrounds from the auto-correlations, we use the ratios of auto-correlations of different tiles to capture the tile to tile variation in the gain bandpass structures. After removing tile-to-tile variations from the gains, we then use the global bandpass average to mitigate calibration errors that introduce contamination in the EoR window, as suggested by \cite{barry2016calibration}. The remainder of this section details the steps of our auto-correlation bandpass calibration technique.

In this section, to distinguish our per frequency calibration from the true gains, we denote the true gain of tile $i$ as $g_i$, and the per frequency calibration solution obtained from section \ref{sec:perfreq} as $\hat{g}_i$.

\subsubsection{Gain Amplitudes}

\begin{figure*}
    \centering
    \includegraphics[width=\linewidth]{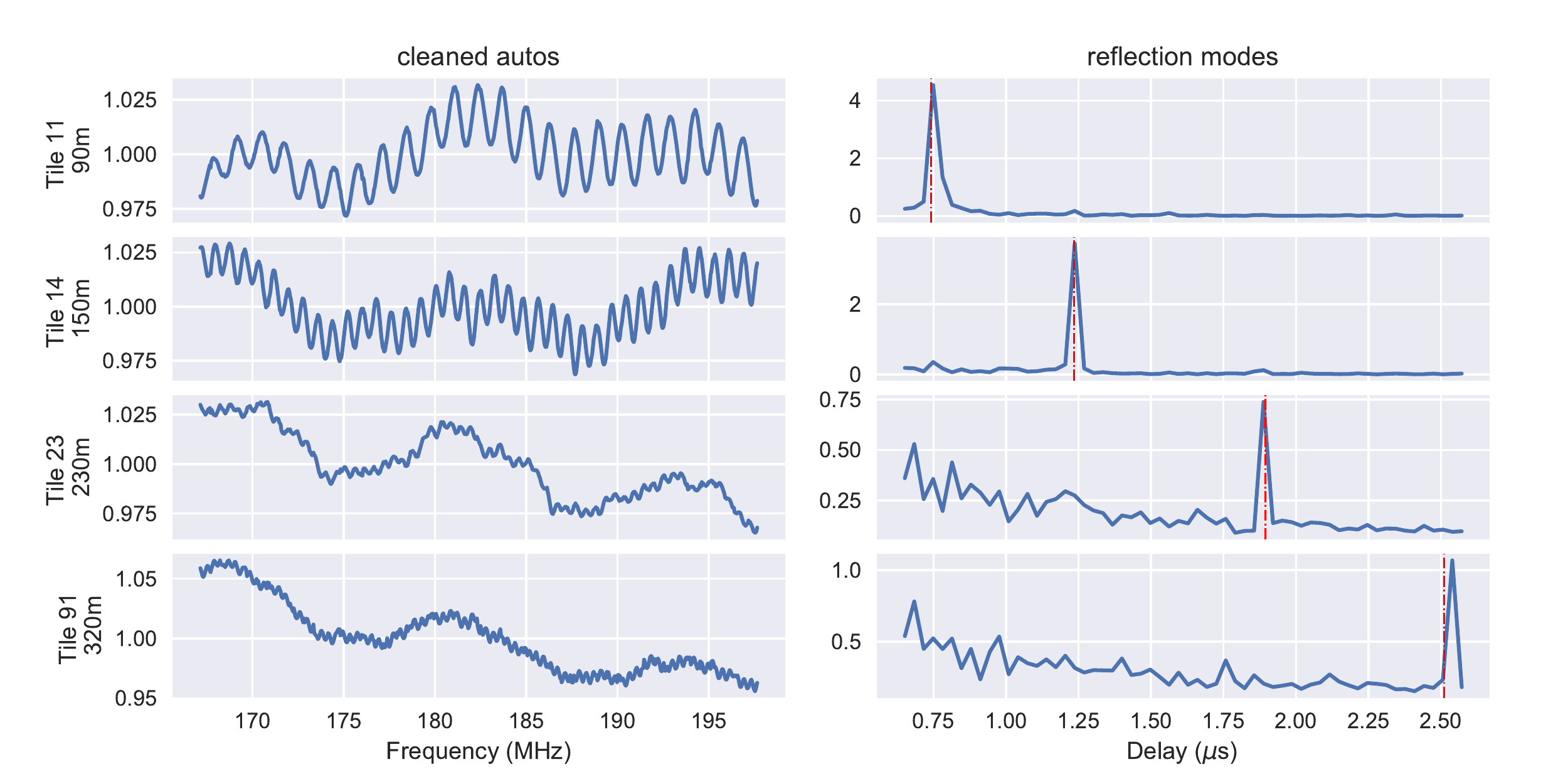}
    \caption{Left column: the cleaned autos for Tile 11 (90m cable), 14 (150m cable), 23 (230m cable) and 91 (320m cable) from top to bottom; Right column: FFT of the left column, vertical dashed red lines mark the theoretical reflection mode. }
    \label{fig:reflection}
\end{figure*}

We begin with the response of a single tile at time $t$ and frequency $f$:
\begin{equation}
    S(t,f) = \int I(\mathbf{l},t,f)B(\mathbf{l},f)\ \mathrm{d}\mathbf{l}+n(t,f),
\end{equation}
where $\mathbf{l}$ is the angular coordinates of the radio sky, $I$ is the sky brightness in Jy, $B$ is the tile beam response, and $n(t,f)\sim N(0,\sigma^2(f))$ is Gaussian random noise. We denote $\mu(t,f)\equiv\int I(\mathbf{l},t,f)B(\mathbf{l},f)$ $\mathrm{d}\mathbf{l}$. Here both $\mu$ and $n$ are real valued. In principle, the auto-correlation of tile $i$ is given by
\begin{equation}
    \begin{split}
&\mathrm{Auto}_i(t,f)=\ |g_i(f)|^2|S(t,f)|^2\\
    &=\ |g_i(f)|^2[\mu^2(t,f)+2\mu(t,f)\bar{n}_i(t,f)+n_i^2(t,f)]
    \end{split}
    \label{eq:autocorr}
\end{equation}
Here we denote the noise in the cross term as $\bar{n}(t,f)$ instead of $n(t,f)$ to distinguish it from the square of the noise. This is because $\mu(t,f)n(t,f)$ has 0 mean and it gets reduced as we average data into a coarse time resolution, which is 2 seconds. However, the self correlation of the noise is a biased term and it does not integrate down over time. To deal with the unknown global sky signal and self correlated noise, we define the `auto ratio' of tile $i$ and tile $j$ as $A^j_i(f)$:
\begin{equation}
\label{eq:auto}
    \begin{split}
    A^j_i(f)\equiv&\biggl<\sqrt{\frac{\mathrm{Auto}_i(t,f)}{\mathrm{Auto}_j(t,f)}}\biggr>_t \\
    =&\frac{|g_i|}{|g_j|}\biggl<\sqrt{\frac{\mu^2+2\mu\bar{n}_i+n_i^2}{\mu^2+2\mu\bar{n}_j+n_j^2}}\biggr>_t
    \end{split}
\end{equation}
where $\langle\rangle_t$ denotes an average over time within an observation snapshot. Assuming all $n_i(t,f)$ are identically distributed as $N(0,\sigma^2(f))$ and $\frac{\sigma(f)}{\mu(t,f)}\ll1$, we keep the first order of $\frac{\bar{n}}{\mu}$ and $\frac{n^2}{\mu^2}$ in Equation \ref{eq:auto}:
\begin{equation}
    \begin{split}
        A^j_i\approx&\frac{|g_i|}{|g_j|}\biggl<\sqrt{(1+\frac{2\bar{n}_i}{\mu}+\frac{n_i^2}{\mu^2})(1-\frac{2\bar{n}_j}{\mu}-\frac{n_j^2}{\mu^2})}\biggr>_t\\
        \approx&\frac{|g_i|}{|g_j|}\biggl<\sqrt{1+\frac{2}{\mu}(\bar{n}_i-\bar{n}_j)+\frac{1}{\mu^2}(n_i^2-n_j^2)}\biggr>_t\\
        \approx&\frac{|g_i|}{|g_j|}\biggl<[1+\frac{1}{\mu}(\bar{n}_i-\bar{n}_j)+\frac{1}{2\mu^2}(n_i^2-n_j^2)]\biggr>_t.
    \end{split}
    \label{eq:autoratio}
\end{equation}
By evaluating Equation \ref{eq:autoratio}, we find 2 advantages of using the auto ratio $A^j_i(f)$ rather than the auto-correlation itself. First, unlike the auto-correlations, $A^j_i(f)$ does not have a noise bias. Second, the auto-correlation is time dependent because the global sky signal $\mu$ is a function of time, while the ratio is time-independent, because the part in $\langle\rangle_t$ is stable over time, and its time variance is $\sim 2.5\times 10^{-5}$. Therefore, the uncertainty of $A^j_i(f)$ is below 0.07\% as the average is calculated over 56 time samples. We can therefore approximate
\begin{equation}
    A^j_i(f)=\frac{|g_i(f)|}{|g_j(f)|},
    \label{eq:ratioapprox}
\end{equation}
which gives the tile to tile variation in the gain amplitudes. Now we have the relative gain amplitudes given by just auto-correlations. The next step is to solve for the absolute gain amplitudes, using the per frequency calibration solutions $\hat{g}_i(f)$ we have in section \ref{sec:perfreq}. Based on equation \ref{eq:ratioapprox}, it is straightforward to see that 
\begin{equation}
    |g_i(f)|\approx A^j_i(f)|\hat{g}_j(f)|,
    \label{eq:aproxgo}
\end{equation}
thus the quantity $A^j_i(f)|\hat{g}_j(f)|$ for every $j$ is an independent realization of $|g_i(f)|$. This gives an opportunity to obtain a smoothed estimation of $|g_i(f)|$ by averaging $A^j_i(f)|\hat{g}_j(f)|$ across all $j$'s: 
\begin{equation}
    |g_i(f)| \approx \biggl<A^j_i(f)|\hat{g}_j(f)|\biggr>_j,
    \label{eq:estgi}
\end{equation}
This incoherently averages down both the noise and the artificial spectral structure (introduced by imperfect sky model and discrepancy in baseline redundancy) in the per frequency calibration, as motivated by \citealt{barry2016calibration}. 

The above process can be applied to every tile $i$, although in practice it is not necessary to repeat the average in equation \ref{eq:estgi} for every tile. Once the amplitude of one tile $i$ has been solved, we can quickly estimate the amplitude of other tiles by multiplying the auto ratio with the average:
\begin{equation}
    |g_k(f)| \approx A^i_k(f)\times \biggl<A^j_i(f)|\hat{g}_j(f)|\biggr>_j,
\end{equation}
as suggested by equation \ref{eq:ratioapprox}.

\subsubsection{Phase}
\label{subsec:cableref}

The phase part of the gain is estimated as:
\begin{equation}
    \arg(g_i(f)) \approx \tau_if+\phi_i+R_i(f),
\end{equation}
where $\tau_i$ and $\phi_i$ are parameters of linear fitting to the phase part of $\hat{g}_i(f)$ over frequency. $\tau_i$ is associated with an tile delay and $\phi_i$ is a phase offset. As there is a small impedance mismatch at the termination of each cable, we also fit a sinusoidal cable reflection term in the phase, denoted by $R_i(f)$. The contribution of cable reflections to the gain amplitudes is already included in auto-correlations, thus we only need to fit for reflection modes in the phase, which we now describe in detail.

We subtract the linear phase from $\arg(\hat{g}_i)$, and define the phase residual as
\begin{equation}
    r_i(f)=\arg(\hat{g}_i(f))-\tau_if-\phi_i.
\end{equation}
We then fit for the reflection term in the phase residual. As there is no absolute phase in interferometry, we pick a tile as the reference tile and force it to have 0 phase. There is no preference for the reference tile selection. For the record, we use Tile 12 as the reference tile. The phase of each of the rest of tiles is the phase relative to the reference tile. Therefore, the phase part of tile $i$ has a hidden additional term, which is the negative phase of the reference tile. This mixes cable reflection modes between tiles. To deal with this challenge, we define a `cleaned phase residual': 
\begin{equation}
    r_i^\prime(f)\equiv\biggl<r_i(f)-r_j(f)\biggr>_{j,c_j\neq c_i},
    \label{eq:phsref}
\end{equation}
where $c_i$ denotes the cable type of tile $i$. We force each term in the average with subscript $j$ to have different cable type from tile $i$, otherwise the same reflection modes from different tiles will be entangled. This average effectively down weights the reflection modes mixed in from other tiles, and simultaneously lowers the noise in the phase of the per frequency calibration by a factor of $\sim$1.4. $r_i^\prime$ now should be ready for cable reflection fitting. 

Instead of using noisy calibration results for reflection mode searching, we again turn to the auto-correlations to better locate the reflection modes. As $A^j_i(f)$ contains bandpass of both tile $i$ and $j$, we need to disentangle them to find the reflection mode for each individual tile. Similar to $r_i^\prime$, we define `cleaned autos' of tile $i$:
\begin{equation}
    A^\prime_i(f)\equiv\biggl<A^j_i(f)\biggr>_{j,c_j\neq c_i}
    \label{eq:autoref}
\end{equation}
We show an example of $A^\prime_i(f)$ for each cable type in Figure \ref{fig:reflection}. The left column shows the `cleaned autos', and the right column shows their Fast Fourier Transform (FFT). The vertical red lines highlight the theoretical reflection mode calculated based on the light travel time in each cable type. This shows a good agreement between the reflection measurement from the data and the theoretical value. We then use the mode we hyperresolved in $A^\prime_i(f)$ to fit $r_i^\prime(f)$.

\subsection{Calibration Summary}
\label{sec:calsum}

A summary of our calibration procedure is as follows.  We implement hybrid calibration by performing an FHD sky-model-based approach first and subsequently apply the redundant calibration algorithm. To mitigate artifacts from imperfect calibration, we further do a bandpass calibration using auto-correlations. We decompose ``auto-correlation ratios" from amplitudes, then do a global bandpass average among all tiles. We fit a linear function for the phase, then fit cable reflection modes to phase residuals using the modes we find in auto-correlations.  In section \ref{sec:tech-comparison}, we will investigate the effects of our new calibration techniques on our power spectrum results.

\section{Power Spectrum Analysis}
\label{sec:psalyz}

Before presenting our power spectrum results in section \ref{sec:res}, this section presents some of the details of our analysis as implemented in our modified FHD/$\varepsilon$ppsilon pipeline, including a description of the regions of $k$ space used in the final result (section \ref{sec:kspace}).  We also present a comparison of our analysis results with those from the RTS/CHIPS pipeline to serve as a further demonstration of the robustness of our analysis (section \ref{sec:rtschips}.)

\subsection{$k$ Space Selection}
\label{sec:kspace}

\begin{figure*}
    \centering
        \begin{subfigure}
        \centering
        \includegraphics[width=0.65\linewidth]{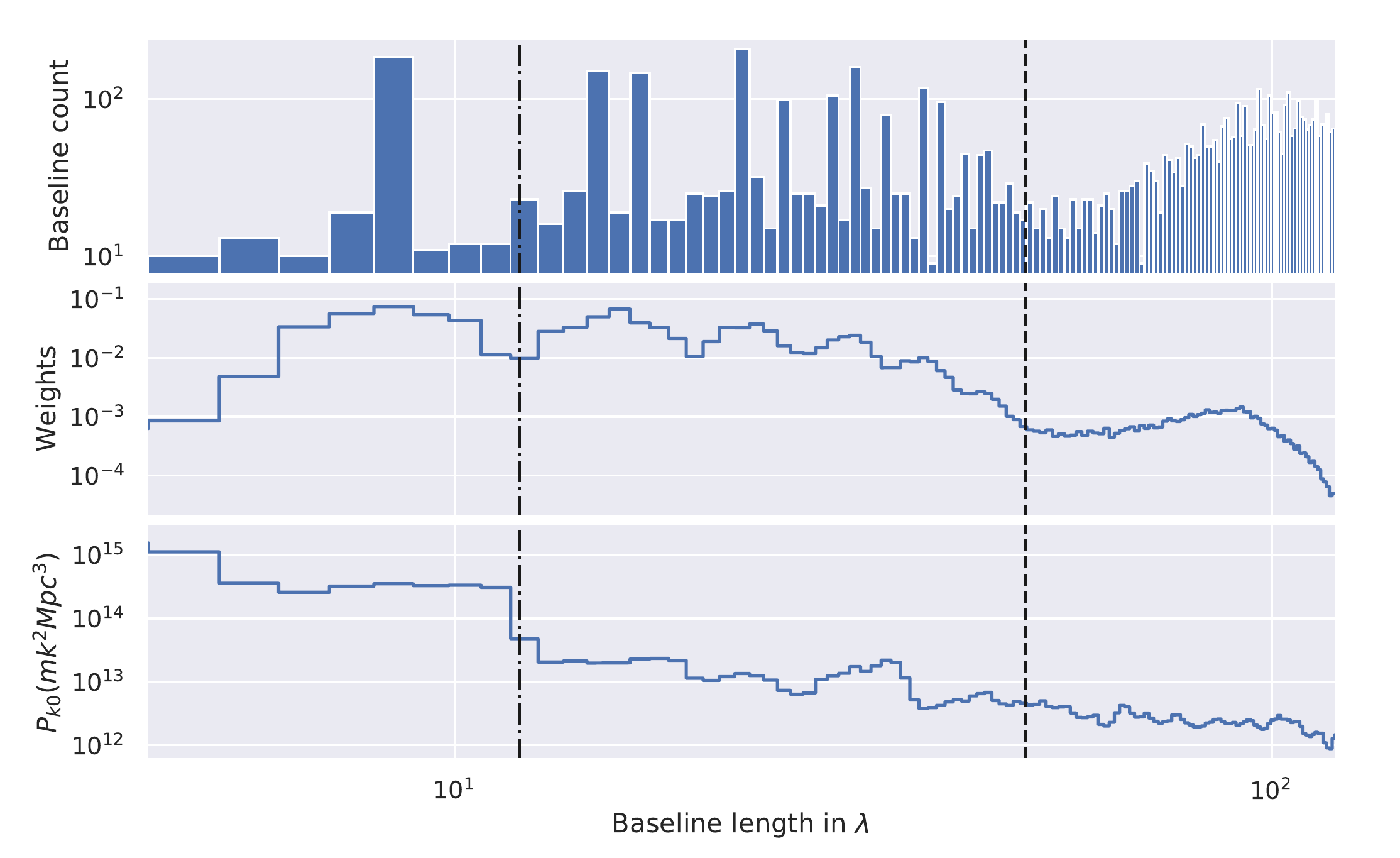}
        \end{subfigure}
        \begin{subfigure}
        \centering
        \includegraphics[width=0.32\linewidth]{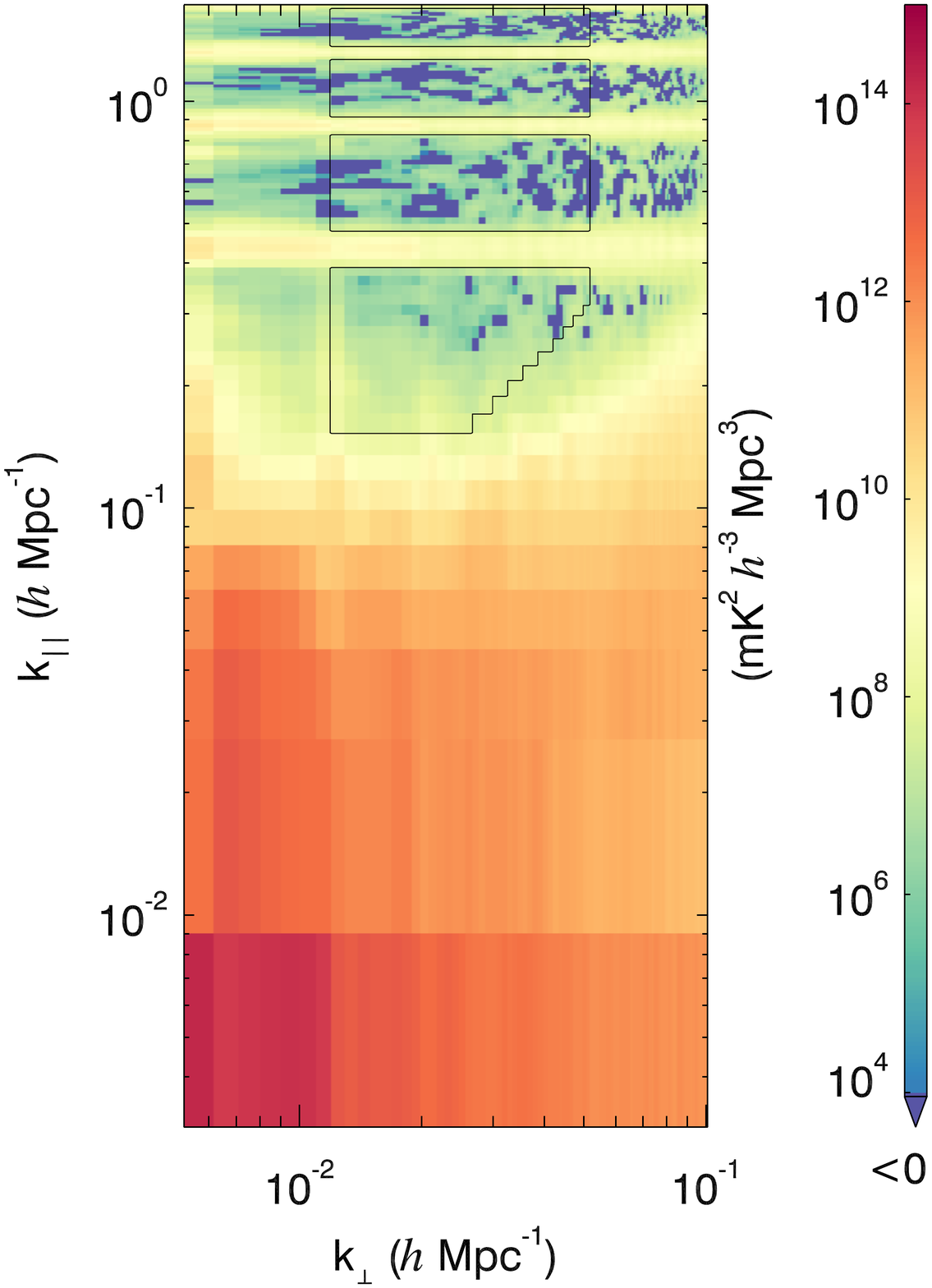}
        \end{subfigure}
    \caption{Left: Baseline cuts in uv space. The vertical dashed line marks the lower bound (12 wavelengths) and upper bound (50 wavelengths) of $k_\perp$ selection for 1d power spectra. Top: Histogram of baseline density of MWA Phase II compact array; Middle: uv weights versus baseline length in wavelengths; Bottom: The power at $k_\parallel$=0; Right: 2d power spectrum for N-S polarization. The contours shows the modes we select for 1d power spectra.}
    \label{fig:pscuts}
\end{figure*}

It is suggested by the cosmological principle that the power spectrum is spherically symmetric \citep{furlanetto2006cosmology,pritchard201221}, thus we can reduce our 3-D $k$ space measurements to a 1-D power spectrum by performing a spherical average along shells of constant $|k|$. To produce the best estimate of the 21 cm power spectrum, we only select modes that are relatively free from foregrounds and coarse band harmonic contamination for inclusion in the spherical average. We emphasize that our mode selection is performed in 3-D $k$ space, but for visual simplicity, we illustrate our selections with the 2-D power spectrum, in which the spatial $k_\perp$ modes have already been averaged over. Our $k$-space selection can be seen in the right panel of Figure \ref{fig:pscuts}: the black contours show the $k$ modes included in the final 1-D power spectrum calculation. We describe the rationale for this mode selection below.

First, we apply the same $k_\parallel$ vs $k_\perp$ wedge slope as \cite{beardsley2016first} and \cite{barry2019newlimit}.
We apply a lower bound of $k_\parallel$ is $0.15$ $h$ $\mathrm{Mpc^{-1}}$ to remove heavily contaminated line-of-sight modes. To avoid coarse band contamination, we avoid 5 $k_\parallel$ bins around the center of each coarse band mode.

We also apply cuts in the $k_\perp$ direction, keeping modes between a lower bound of $k_\perp =  12\lambda$ and an upper bound of $50\lambda$, where $\lambda$ denotes wavelength. The left panel of Figure \ref{fig:pscuts} shows the justification for these cuts. The top panel shows the histogram of baseline density. The overwhelmingly large spikes in the histogram are contributed by the highly redundant baselines. The middle panel shows the weights, i.e., the relative number of measurements at each $k_\perp$ bin. The weights are equivalent to convolving the top histogram with our modified gridding kernel (see section \ref{sec:pipe-details}). The bottom plot shows the 1-D power at $k_\parallel=0$, which is dominated by foregrounds. 

As we can see in the bottom plot of the left panel, as well as the bottom slice of the 2-D power spectrum on the right, the foregrounds are extremely strong $k_\parallel$ below $12\lambda$. There is a significant drop of foregrounds at $12\lambda$ which is marked by vertical dot-dashed line in Figure \ref{fig:pscuts}. We can also see in the 2-D power spectrum that foregrounds at $k_\perp<12\lambda$ are coupled to high $k_\parallel$ modes. Thus we drop $k_\perp$ modes below $12\lambda$. This cut does remove the 14 m redundant baselines, but the second shortest redundant baselines are still included in our selection, so we do not lose significant power spectrum sensitivity. 

The upper bound of the $k_\perp$ selection is based on the uv weights. The weights drop to a local minimum around $50\lambda-60\lambda$, so increasing the upper bound does not significantly increase signal to noise. However, as the 2-D power spectrum shows, at higher $k_\perp$ the width of the coarse band harmonic contamination gets wider, affecting more $k_\parallel$ modes near the coarse band modes. We thus exclude $k_\perp$ modes greater than $50\lambda$, as marked by vertical dashed line in Figure \ref{fig:pscuts}.

\subsection{RTS/CHIPS Comparison}
\label{sec:rtschips}

\begin{figure*}
    \centering
    \includegraphics[width=\linewidth]{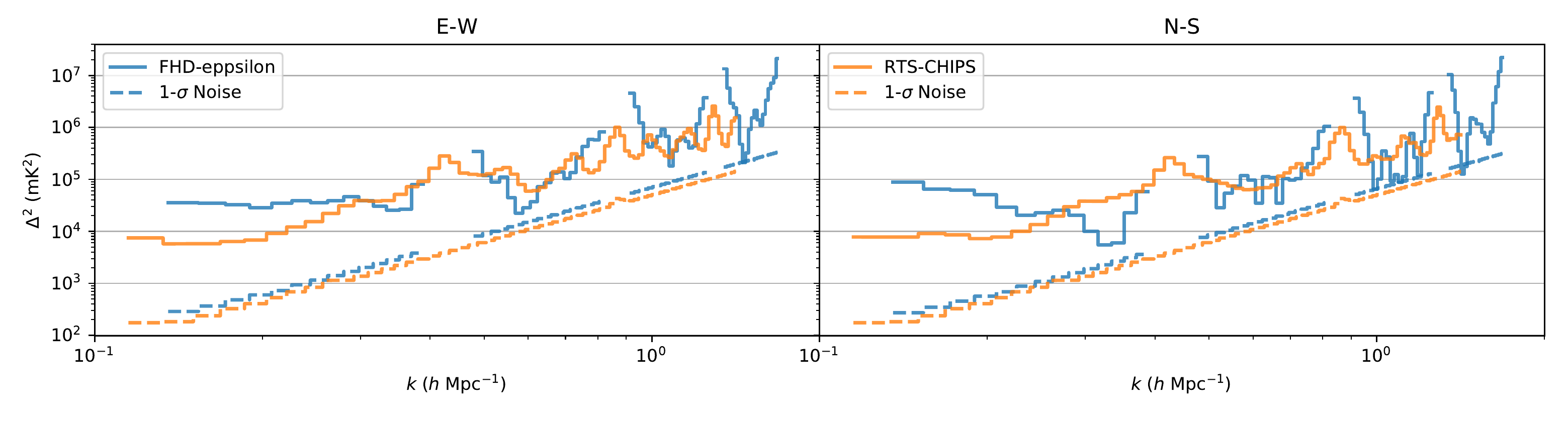}
    \caption{Comparison of $2-\sigma$ limits on the 1-D power spectrum between the FHD/$\varepsilon$ppsilon pipeline (blue) and RTS/CHIPS (orange) pipeline using the wedge cuts shown in Figure \ref{fig:pscuts}.  Dashed lines show the $1-\sigma$ noise levels calculated by the pipelines. The E-W polarization is on the left and the N-S polarization is on the right.}
    \label{fig:rtsvsfhd}
\end{figure*}

\cite{jacobs2016murchison} has demonstrated the value of comparing two independent analysis pipelines for validating an analysis. As a robustness check, we compare the FHD/$\varepsilon$ppsilon pipeline with the Real Time System (RTS, \cite{mitchell2008real,ord2010interferometric}) and Cosmological H I Power Spectrum (CHIPS, \cite{trott2016chips}) pipeline on a subset of our data set. The RTS plays a similar role to FHD, but they are methodologically different \citep{jacobs2016murchison}. RTS performs direction dependent gain calibration using a peeling method with many calibrator sources in the radio sky, and also performs an ionosphere refraction correction. This RTS calibration procedure is fundamentally different from the FHD + omnical + auto correlation bandpass fitting technique in this work, thus it is a good validation for our calibration step. 

The visibilities processed by RTS are the input of the CHIPS power spectrum pipeline. Unlike $\varepsilon$ppsilon, CHIPS applies $w$-projection to account for sky curvature and directly grids data in $(u,v,f)$ space rather than transforming to image space. CHIPS also applies a kriging step to model the missing channels, and an optimal Fourier Transform to give the best power spectrum estimation. 

For this comparison, we choose to analyze a subset of our data, which are selected 4 hours of data from zenith pointing observations. 
We believe the zenith observations to have the best sensitivity and least contamination from foregrounds among the 5 pointings \citep{barry2019newlimit}. 
We calculate the power spectrum using the whole 30.72 MHz band with a center frequency of 182.395 MHz. Because we apply a Blackman-Harris window function in the frequency transform, the effective bandwidth is 15.36MHz; this range of frequencies is too wide to have a simple cosmological interpretation, but is useful for increasing our $k$-space resolution in the analysis comparison. The comparison of the $2-\sigma$ power spectrum upper limits from both pipelines is shown in Figure \ref{fig:rtsvsfhd}. 
The results from the two pipelines are generally consistent, although some interesting difference can be seen.  First, it appears that RTS/CHIPS better suppresses foreground power in the lowest $k$ modes of the power spectrum.  CHIPS's kriging technique (which interpolates over missing channels using Gaussian process regression) can also mitigate the effects of the missing channels at coarse band edges. We flag the $k$ modes most contaminated by the coarse band harmonics in FHD/$\varepsilon$ppsilon, but even outside these flagged areas the RTS/CHIPS analysis shows lower power.  In the areas between the coarse band harmonics, however, FHD/$\varepsilon$ppsilon generally has lower power, potentially due to our modified gridding kernel, calibration techniques, or the better removal of contaminated data through our quality metrics.  Lastly, the estimated noise level of the RTS/CHIPS analysis is $\sim30\%$ lower than that from FHD/$\varepsilon$ppsilon.  While it may seem surprising to have different noise levels estimated from the same data set, a disagreement at this level might be expected.  As described in section \ref{sec:pipe-details}, we use a modified kernel in our gridding process, which can suppress systematic errors but decreases the signal-to-noise (SNR) by $\sim25\%$ compared to the optimal value \citep{morales2009software,barry2019newlimit}; CHIPS also uses a modified gridding kernel, but one distinct from that used here, in that it was chosen to have similar SNR to the optimal value.

\section{Results}
\label{sec:res}
\begin{figure*}
    \centering
    \begin{subfigure}
        \centering
        \includegraphics[width=0.32\linewidth]{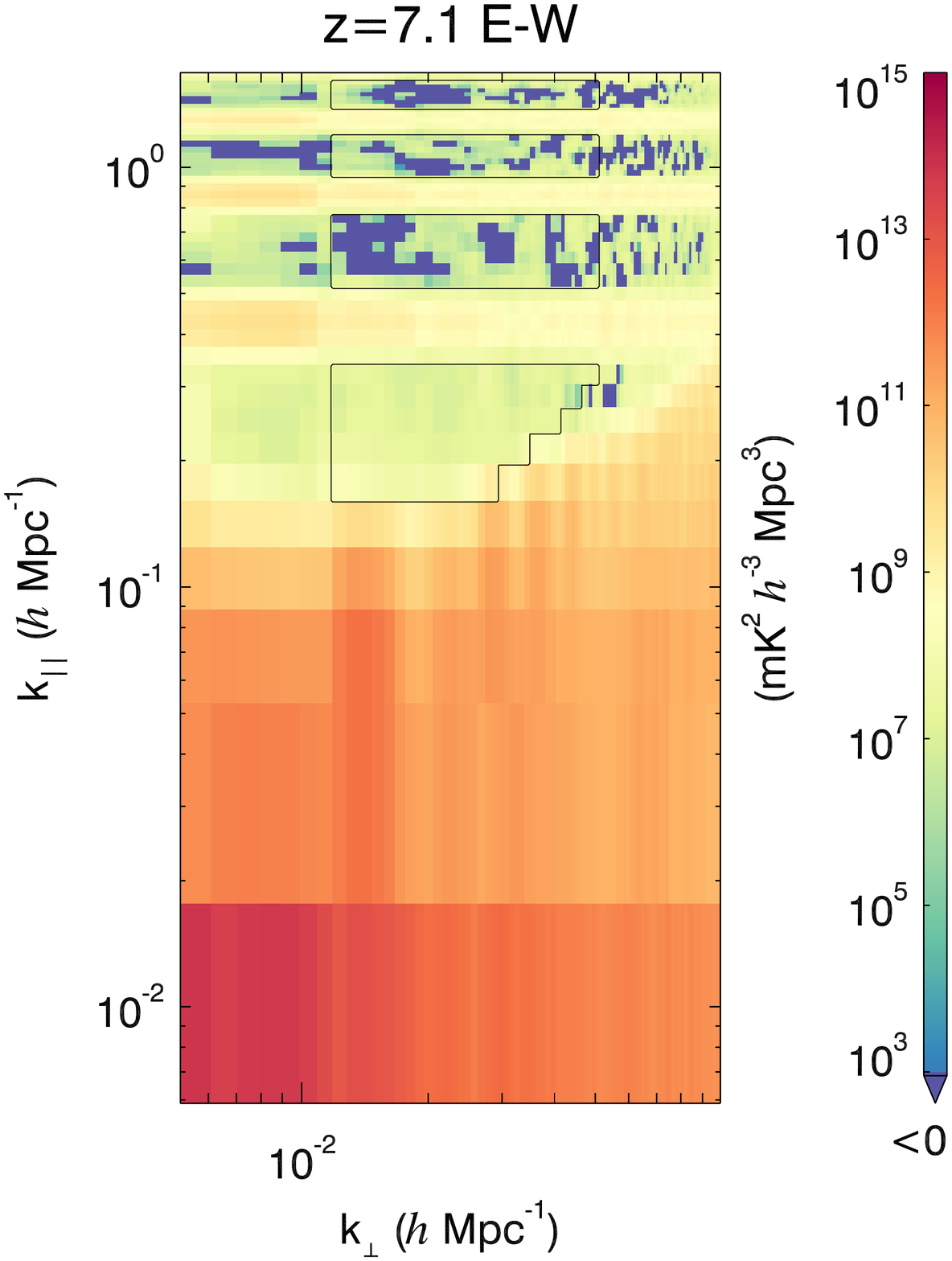}
    \end{subfigure}
    \begin{subfigure}
        \centering
        \includegraphics[width=0.32\linewidth]{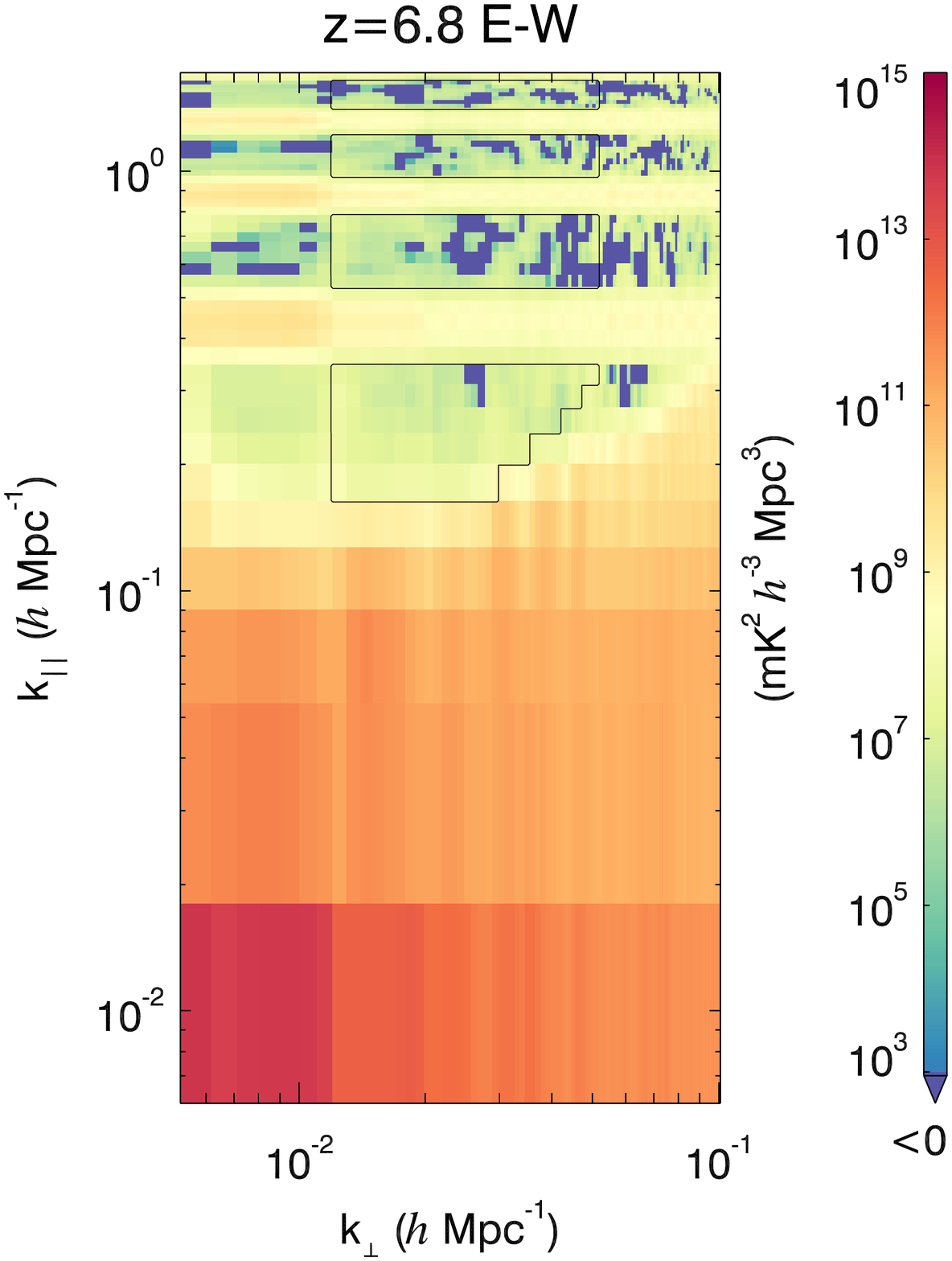}
    \end{subfigure}
    \begin{subfigure}
        \centering
        \includegraphics[width=0.32\linewidth]{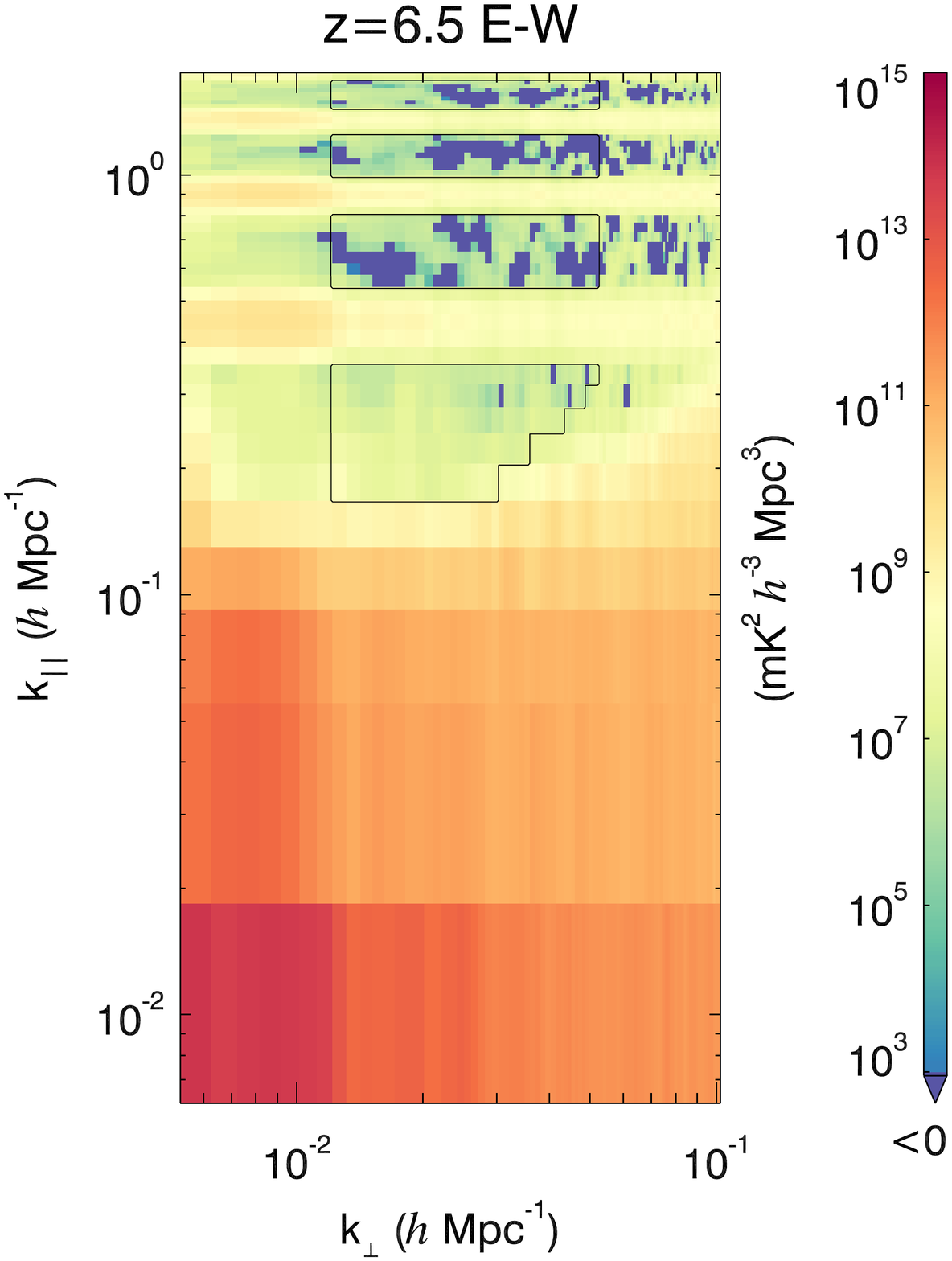}
    \end{subfigure}
    \begin{subfigure}
        \centering
        \includegraphics[width=0.32\linewidth]{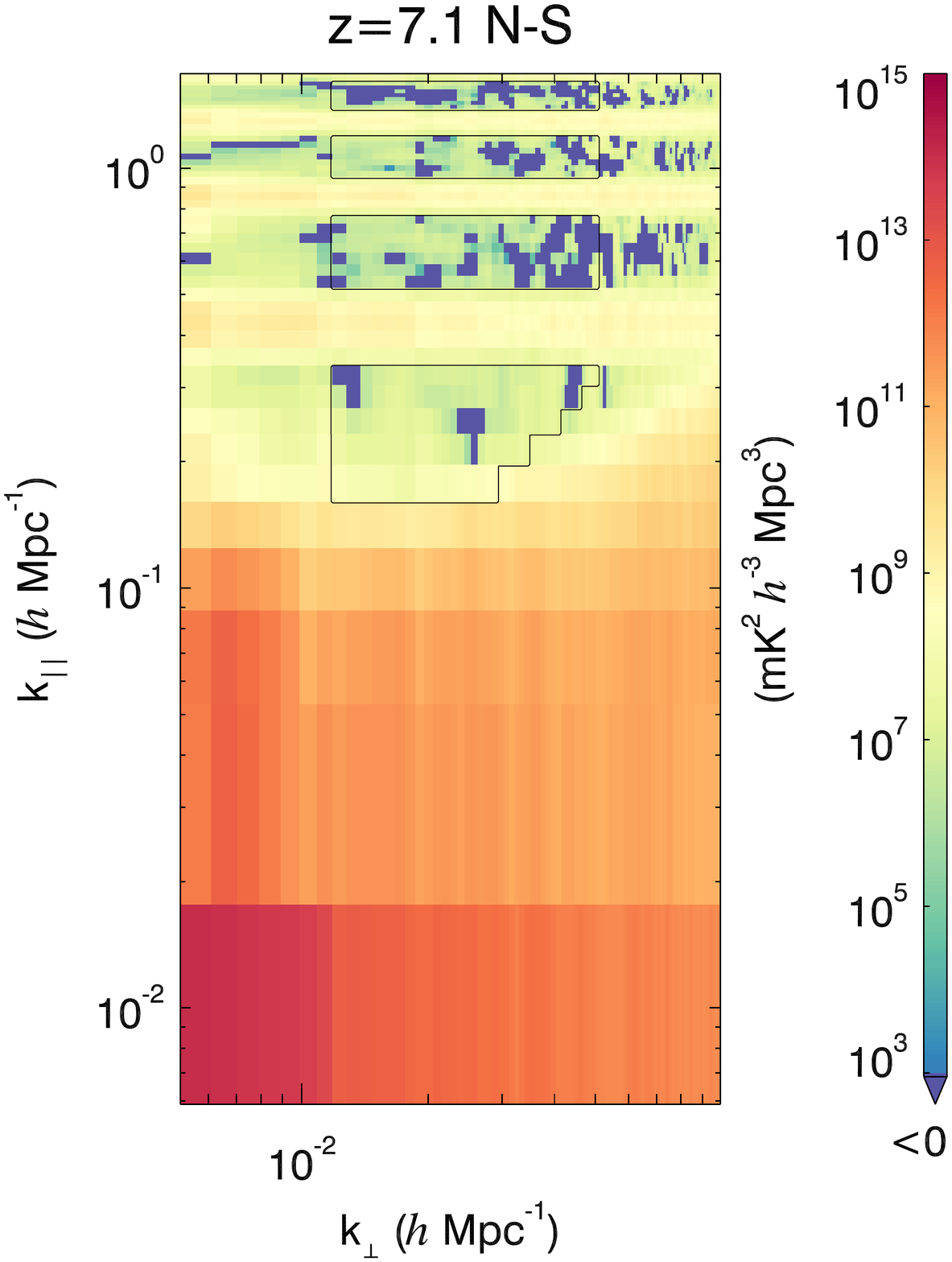}
    \end{subfigure}
    \begin{subfigure}
        \centering
        \includegraphics[width=0.32\linewidth]{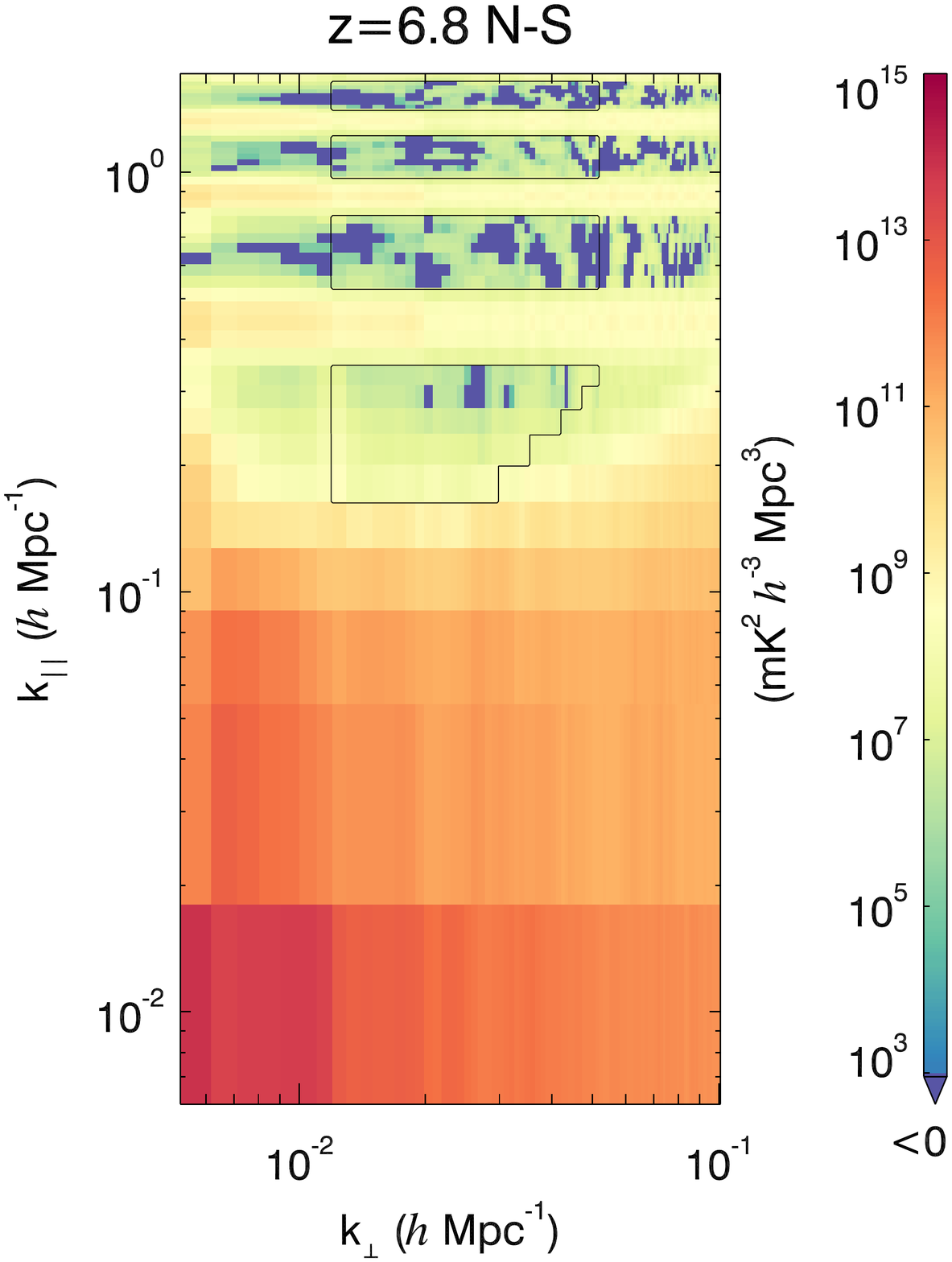}
    \end{subfigure}
    \begin{subfigure}
        \centering
        \includegraphics[width=0.32\linewidth]{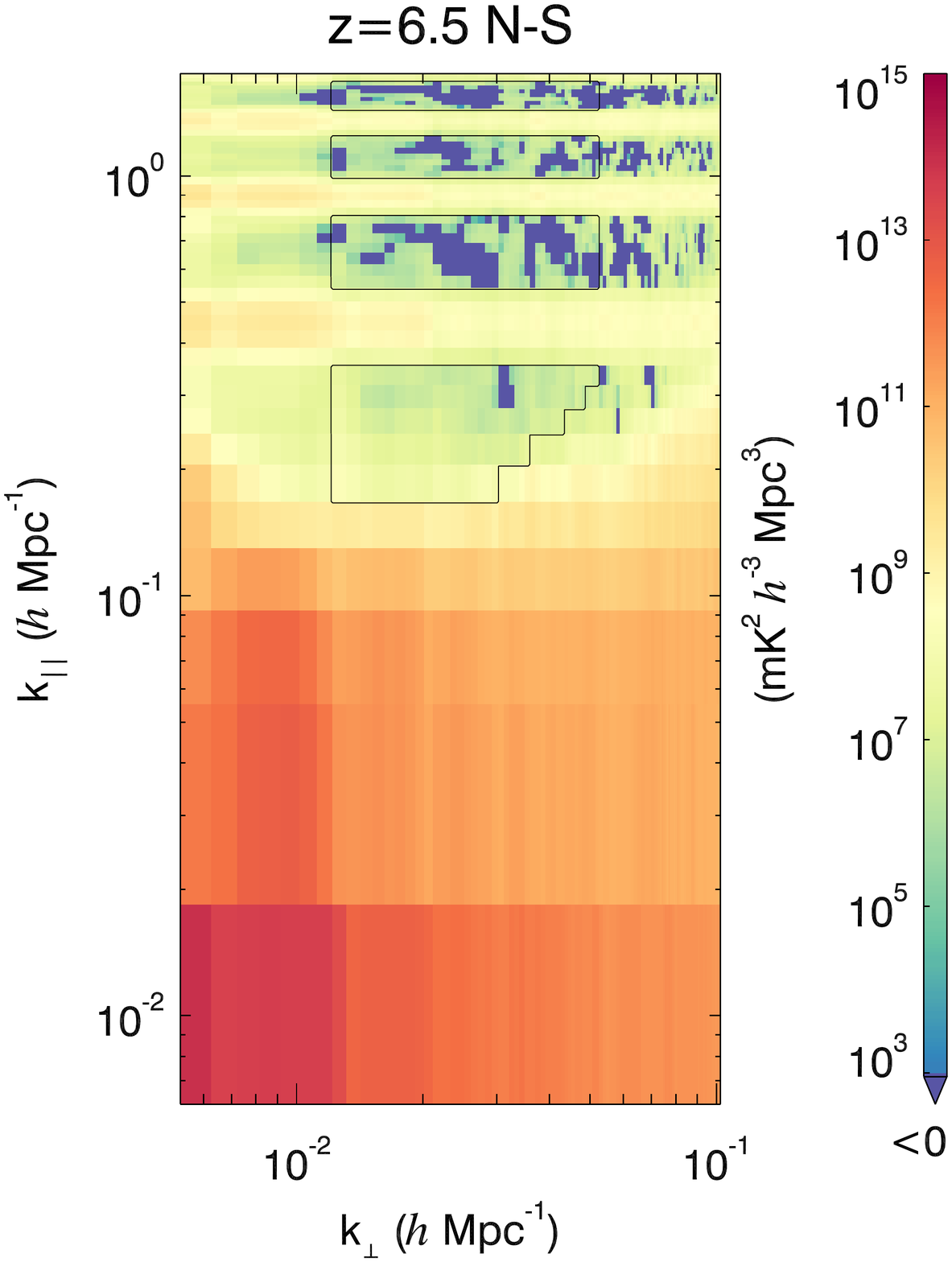}
    \end{subfigure}
    \caption{2-D power spectra for EoR power spectrum limit at redshift 7.1 (left column), 6.8 (middle column) and 6.5 (right column) in East-West polarization (top row) and North-South polarization (bottom row). The black contours illustrate the modes we select for the 1-D power spectra calculation. }
    \label{fig:2dlimit}
\end{figure*}

\begin{figure*}
    \centering
    \includegraphics[width=\linewidth]{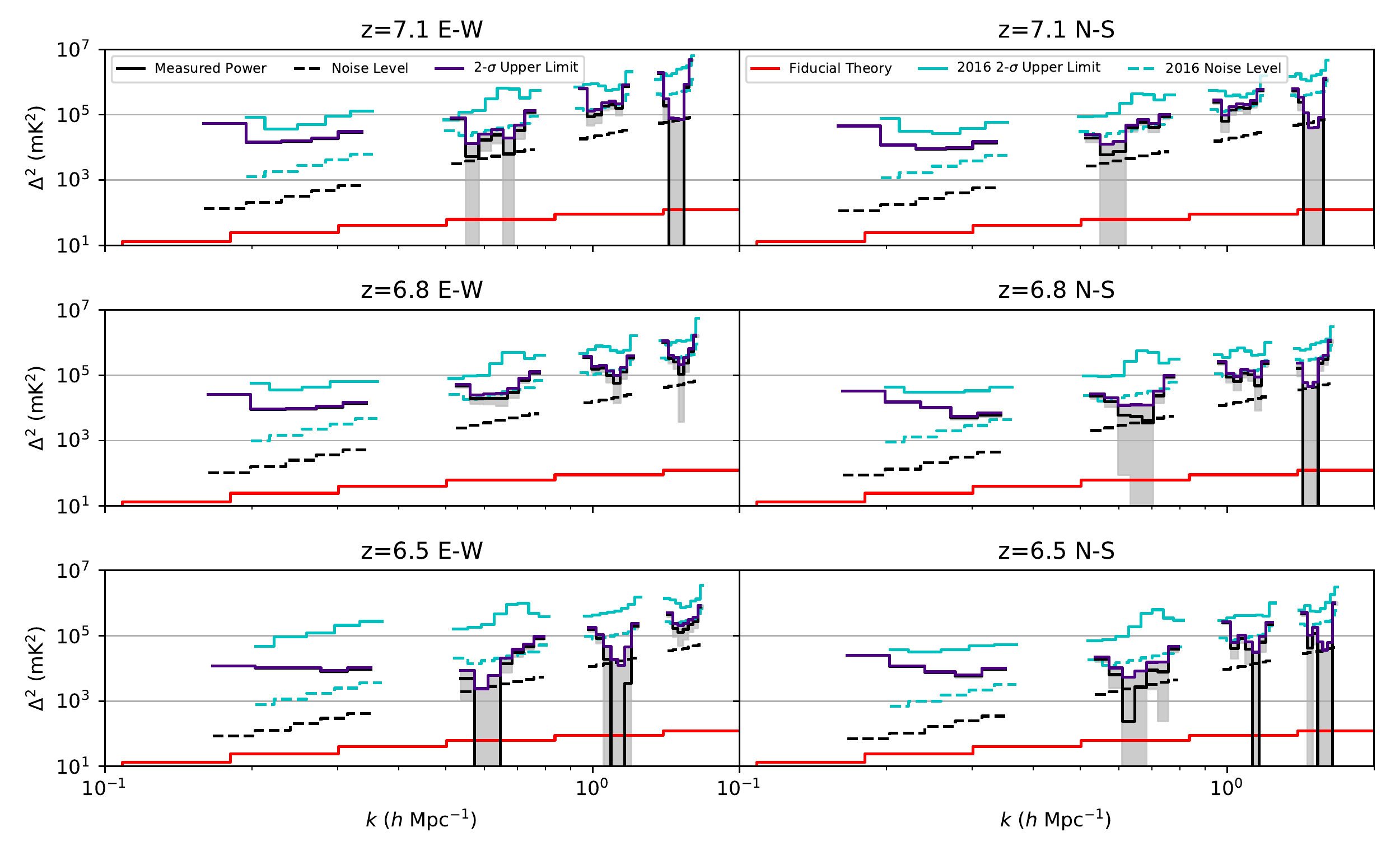}
    \caption{Power spectrum upper limit for redshift 7.1 (top), 6.8 (middle), 6.5 (bottom) in E-W polarization (left) and N-S polarization (right). The black solid line shows the measured 1d power and the purple line shows the corresponding 2-$\sigma$ power spectrum upper limit. The black dashed line shows the noise level. The gray boxes show the $\pm 2\sigma$ error bar. The cyan solid line and cyan dashed line show the 2-$\sigma$ power spectrum upper limit and noise level from \cite{beardsley2016first}, respectively. The red line represents a fiducial EoR level.}
    \label{fig:uplim}
\end{figure*}

\begin{table}
\centering
\caption{Lowest Measured Power Spectrum Values}
\begin{tabular}{|| c c c c ||}
 \hline
 $z$ & $\Delta_\mathrm{UL}^2$ ($\mathrm{mK^2}$) & $\sigma$ ($\mathrm{mK^2}$) & $k$ ($h$ $\mathrm{Mpc^{-1}}$) \\
 \hline
 \hline
 $7.1$ & $9.28\times 10^3$ & $2.76\times 10^2$ & $0.25$ \\ 
 \hline
 $6.8$ & $5.59\times 10^3$ & $3.10\times 10^2$ & $0.29$ \\
 \hline
 $6.5$ & $2.39\times 10^3$ & $2.35\times 10^3$ & $0.59$ \\
 \hline
\end{tabular}
 \label{tb:bestlim}
\end{table}

In this section, we give the final result of this analysis. We calculate the power spectra at redshift 7.1 (174.7 MHz), 6.8 (182.4 MHz) and 6.5 (190.1 MHz), using a bandwith of 15.36 MHz for each redshift. With a Blackman-Harris window applied, the effective bandwidth is 7.68 MHz. Specifically, we make power spectra using sub-bands of $167.1-182.3$ MHz for redshift 7.1, $174.8-190.0$ MHz for redshift 6.8, and $182.5-197.7$ MHz for redshift 6.5.

\subsection{Power Spectrum Upper Limits}

The 2-D power spectra for these 3 sub-bands are shown in Figure \ref{fig:2dlimit}. We draw contours to highlight modes we use for 1-D spherical averaged power spectrum. The 1-D power spectra are shown in Figure \ref{fig:uplim}. The solid black line represents the measured power; dashed black line is the noise level; the grey boxes show $2-\sigma$ error bars; and the purple line is our final $2-\sigma$ upper limit, which corresponds to a 97.7\% confidence interval. The red line is a theoretical EoR level for reference.  The cyan lines are results from \cite{beardsley2016first}, where the solid line shows the $2-\sigma$ upper limit, and dashed line represents the noise level.  (We also present a comparison with \citealt {barry2019newlimit} in Figure \ref{fig:p1p2}, which we discuss below; since they use a band of 168.6 to 187.3 MHz for their analysis, it cannot be directly compared with our three sub-bands.) Our best measurement at each redshift are listed in Table \ref{tb:bestlim} along with the corresponding $k$ modes and 1-$\sigma$ error. $\Delta^2_\mathrm{UL}$ denotes the $2-\sigma$ upper limit (see appendix \ref{apdx:limcal} for the upper limit calculation). The measurement at $k=0.59h$ $\mathrm{Mpc^{-1}}$ and $z=6.5$ is noise dominated, suggesting that we will be able to further improve the limit by integrating
more data.

Note that we interpret all measurements to be upper limits on the 21\,cm power spectrum, even though many of the modes measured are statistically significant detections of power. This is equivalent to identifying all the power we detect as the result of foreground contamination.  At the level of the current measurements, this should not be a particularly contentious interpretation---even our lowest limit of $\Delta^2_{UL} = 2.35 \times 10^3 \mathrm{mK}^2$ is more than an order of magnitude above any standard signal models \citep{pober2014next}.\footnote{Non-standard models used to explain the large signal reported by \cite{bowman2018edges} may cause a boost in the power spectrum amplitude as well, but these effects are not expected to be significant at $z\sim6$ \citep{fialkov2018cosmicdawn,munoz2019vao}.}  Several other lines of evidence in our analysis also suggest that these detections are not cosmological.  First, we can point to many of the steps in our analysis that did remove power from higher $k_{\parallel}$ modes:
the introduction of the modified gridding kernel (section \ref{sec:pipe-details}), the flagging of ultrafaint RFI or other bad data (section \ref{sec:qltm}), and the auto-correlation bandpass calibration (section \ref{sec:bpauto}).  Since all of these techniques reduce power in the measured power spectrum, but do not bias the recovered EoR signal in simulation (section \ref{sec:unbiased}), it seems reasonable to conclude that our detections are caused by further systematic errors that our analysis has not yet removed.  Mis-calibration, in particular, can leave foreground residuals that bias the power spectrum and we already know of further improvements that we could make to our approach (e.g. including a diffuse sky model to better calibrate the shortest baselines.) 

Second, we note that the contaminants in our data do have a strong dependence on the tile beam pointing, which would not be expected from a cosmological signal (Figure \ref{fig:wp}).  The worst pointings are excluded from the analysis entirely, but overall these results provide evidence for residual signals in our data that do not behave like a cosmological signal.  Lastly, we again highlight that the footprint of the detections in 2-D $k$-space (Figure \ref{fig:2dlimit}) have the shape (e.g. the wedge and coarse band lines) that we theoretically expect for foreground residuals in the presence of calibration errors \citep{morales2018understanding}.

Were our detections and limits an order of magnitude lower, however, these arguments would not be sufficient.  The question of validating (or invalidating) the first purported EoR signal detection is one facing every team working in this field.  \cite{pober2016upperlimits} suggest a number of possible tests that could be used.  The most compelling confirmation would likely come from an independent experiment; consistent results from independent pipelines or independent fields on the sky would also bolster any conclusions.  As this work focuses on one field from one experiment, such comparisons are largely outside our present scope.  We do compare with the RTS/CHIPS pipeline (Figure \ref{fig:rtsvsfhd}), which sees similar (but not identical) detections of power in most $k$ modes.  Several null tests are also used as diagnostics in the FHD/$\varepsilon$ppsilon pipeline, including the even/odd power spectrum difference made by splitting the data into two sets on a two second cadence \citep{barry2019fhd}.  This particular test is passed, meaning the result is consistent with the expected noise level in the data, but jackknives over pointing, lst, baseline, and frequency may all play a role in validating future, lower limits from the MWA. 

\section{Comparison with MWA Phase I}
\label{sec:ps-comparison}

\begin{figure*}
    \centering
    \includegraphics[width=\linewidth]{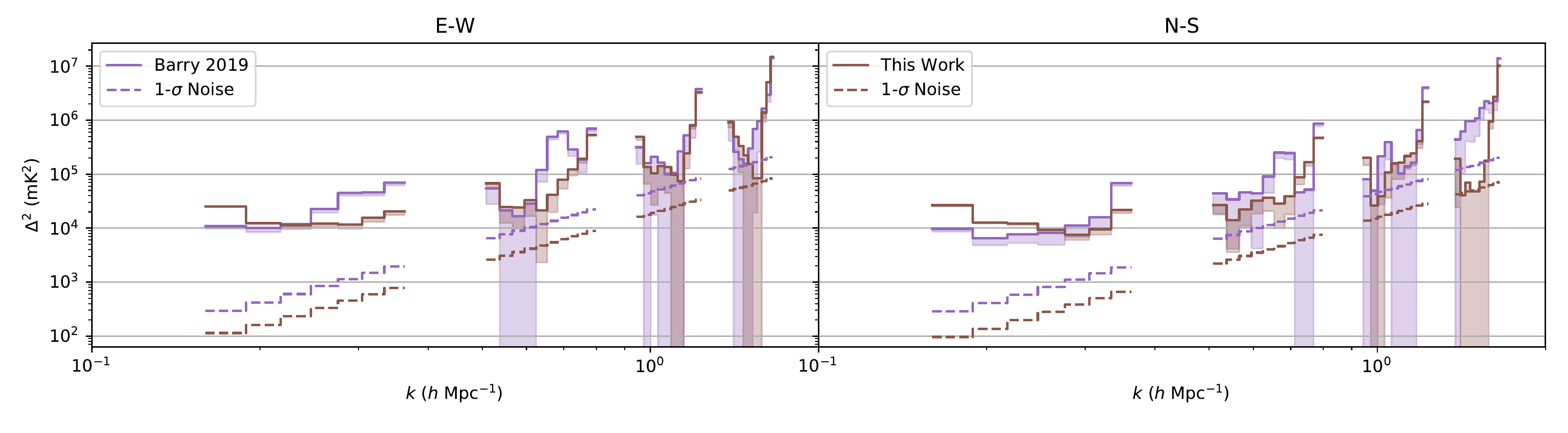}
    \caption{Comparison between Phase I (purple) and Phase II (brown) power spectrum upper limit using the same FHD-$\varepsilon$ppsilon pipeline. The solid lines are 2-$\sigma$ upper limit, dashed lines are 1-$\sigma$ thermal noise.}
    \label{fig:p1p2}
\end{figure*}

It is instructive to compare our results with the two principal limits to come out of Phase I of the MWA: those of \cite{beardsley2016first} and those of \cite{barry2019newlimit}.
As shown in Figure \ref{fig:uplim}, we have improved $\Delta^2$ by about an order of magnitude over \cite{beardsley2016first} at nearly all values of $k$. One striking difference is that the noise curve in our work is 1 order of magnitude smaller. This may seem counter-intuitive, given that \cite{beardsley2016first} has integrated 32 hours of data, while in this work, we have only 19 hours for East-West polarization and 23 hours for North-South polarization. There are, however, two reasons for our lower noise level. First, MWA Phase II added a large number of redundant baselines for the purpose of increasing power spectrum sensitivity. In the modes we select for the 1-D power spectrum, there are many more baselines than for Phase I. Second, in the \cite{beardsley2016first} analysis, to avoid foregrounds outside the main field of a view, only a small area of the sky in the main lobe of the image cube was used for power spectrum analysis. This is equivalent to applying a top-hat window function in image space. This window function introduced systematics to the power spectrum, resulting in an effective increase in the noise level \citep{barry2019newlimit}.

\cite{barry2019newlimit} presents a re-analysis of the data set used in \cite{beardsley2016first} which also improves on that limit by almost an order of magnitude.  In many ways, the \cite{barry2019newlimit} result is a more relevant comparison for understanding the differences between the Phase I and II arrays, since we use the same stage of the FHD/$\varepsilon$ppsilon pipeline and  have both applied SSINS for data quality control.
That work uses 21 hours of data (selected from the 32 hour set that was analyzed in \citealt{beardsley2016first}), about the same amount of data used in this work. \cite{barry2019newlimit} selects frequency range 168.6 MHz - 187.3 MHz, corresponding to a single power spectrum at redshift 7. The motivation of this selection is to avoid a digital gain discontinuity at 187.5 MHz of Phase I. Although the Phase II upgrade has removed the digital gain discontinuity (enabling us to make the three-redshift limits shown in Figure \ref{fig:uplim}), we can use the same frequency range to make a fair comparison with the \cite{barry2019newlimit} result. The 2-$\sigma$ power spectra limit (solid) and 1-$\sigma$ noise curve (dashed) for both works is shown in Figure \ref{fig:p1p2}. The shaded region represents the 2-$\sigma$ error bar. With a similar amount of data being used and same image window kernel being applied, Phase II shows a lower noise curve. This indicates the improved sensitivity of the Phase II design. 

One feature in Figure \ref{fig:p1p2} worth calling attention to is the lack of any noise dominated measurements between the first and second coarse band lines in the Phase II power spectra.
Figure \ref{fig:uplim} indicates that there are multiple noise dominated
bins in this range, except
East-West polarization at $z = 6.8$.  
This discrepancy can be explained
by noting the frequency range that we use
for Figure \ref{fig:p1p2}, which consists of 69\% of the band we use
for $z = 7.1$ power spectra and 53\% of the band we use
for $z = 6.8$. As in Figure \ref{fig:uplim}, the East-West polarization power spectra is highly systematic dominated at
$z = 6.8$, and the noise dominated bins in North-South
polarization at $z = 7.1$ and $z = 6.8$ do not overlap,
therefore the systematic dominated bins overwhelm the
noise dominated bins. This illustrates the value in removing the digital gain discontinuity that was present in Phase I observations and enabling the use of the entire MWA bandwidth.

\subsection{Technique Comparison}
\label{sec:tech-comparison}
Beyond the layout and sensitivity differences between Phase I and Phase II and the different frequency ranges analyzed, the other difference between our work and that of \cite{barry2019newlimit} lies in the calibration.  In this section, we investigate the effects each of our calibration changes had on results in order to better understand their impact.  In particular, we look at the effects each technique had on the power spectrum limits derived from a 5 hour zenith-pointing-only set.  In 3D $k$-space, we subtract power spectra made with/without a particular technique from a fiducial power spectrum and then bin to 1D to better illustrate the change.

\begin{figure*}
     \centering
     \includegraphics[width=\linewidth]{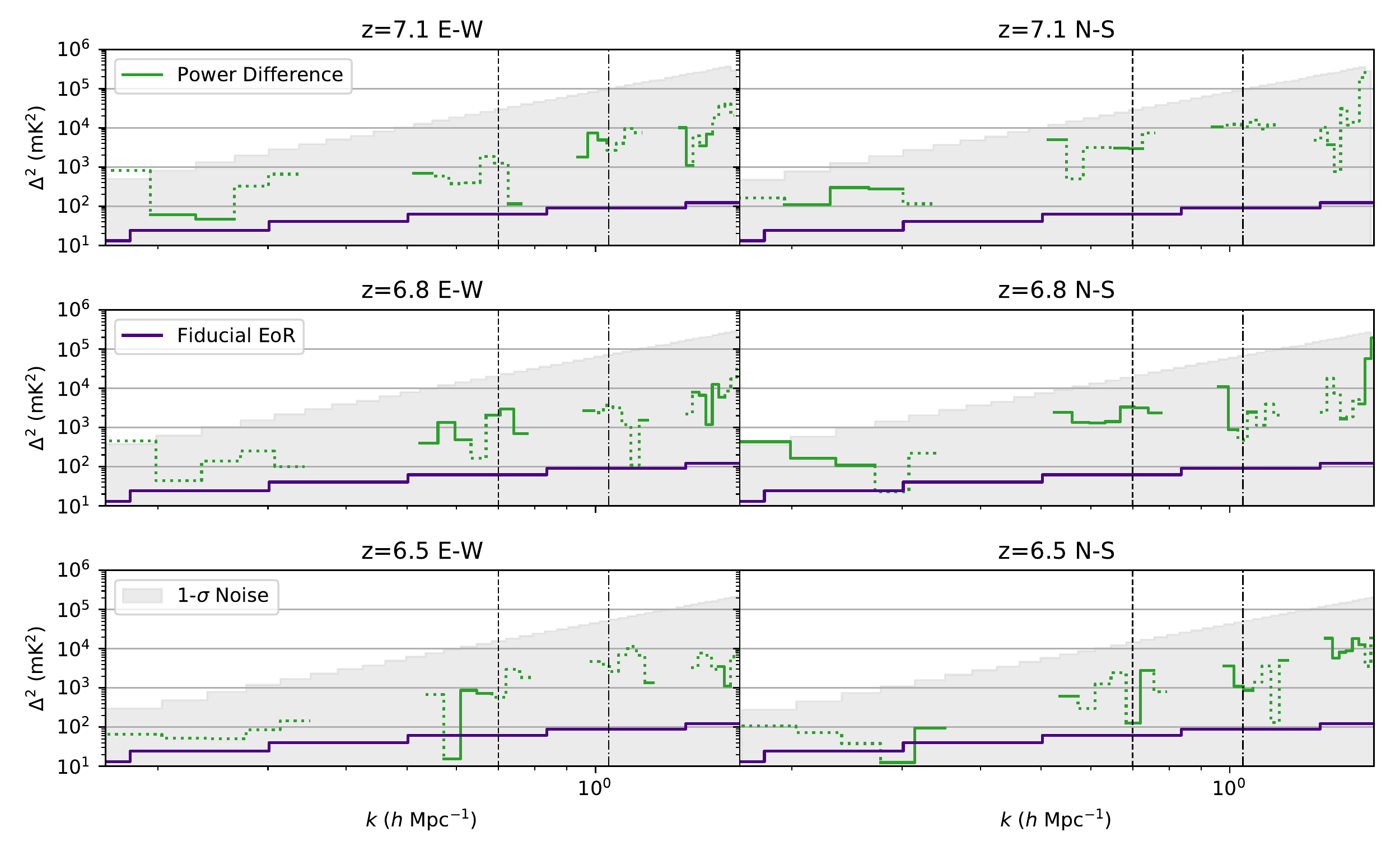}
     \caption{Contributions to power spectrum improvements at East-West polarization (left) and North-South polarization (right) in each of our three redshift bins from hybrid calibration (green). Solid lines illustrate power mitigation and dotted lines indicate an excess of power introduced by hybrid calibration. The shaded region quotes the 1-$\sigma$ noise level. The dark purple line shows a theoretical EoR level for scale. The vertical dashed line marks the 150 m cable reflection mode, and the vertical dashed dotted line marks the 230 m cable reflection mode.}
     \label{fig:cal_diff_hybrid}
 \end{figure*}
First, we can investigate the effect of hybrid calibration by re-running our analysis using only sky-based and bandpass calibration (i.e. excluding the redundant calibration step).  Figure \ref{fig:cal_diff_hybrid} shows power spectrum difference plots made in each of our three redshift bins for each polarization.  The power spectra made without redundant calibration have been subtracted from power spectra made using our fiducial calibration technique (i.e. the technique used to make the limits shown in Figure \ref{fig:uplim}).  Modes with solid lines are positive, indicating that power contamination is mitigated by redundant calibration, while dotted lines are negative, indicating modes where an excess of power has been introduced. The dark purple line marks the theoretical EoR level. The shaded regions indicate the noise level in this data set. We stress, however, that we are subtracting the same data set with different calibrations applied, thus the same noise realization is present in both power spectra.  Changes below the noise level are therefore real changes introduced by the calibration, but the noise level shows an interesting scale for accessing the impact of these changes on our power spectrum limits.  Overall, we see that the changes introduced by redundant calibration do not yield significant improvements; the changes are above the EoR level, but there is little consistency in which modes are improved and which modes see increased contamination.  \cite{li2018comparing}, however, found that redundant calibration yielded a small but consistent improvement in MWA Phase II power spectra.  A key difference between that work and this one, however, is that that work did not include the auto-correlation bandpass calibration we use here.  The results shown in Figure \ref{fig:cal_diff_hybrid} therefore suggest that the small benefits of hybrid calibration shown in \cite{li2018comparing} can also be realized through other improved calibration techniques.

\begin{figure*}
     \centering
     \includegraphics[width=\linewidth]{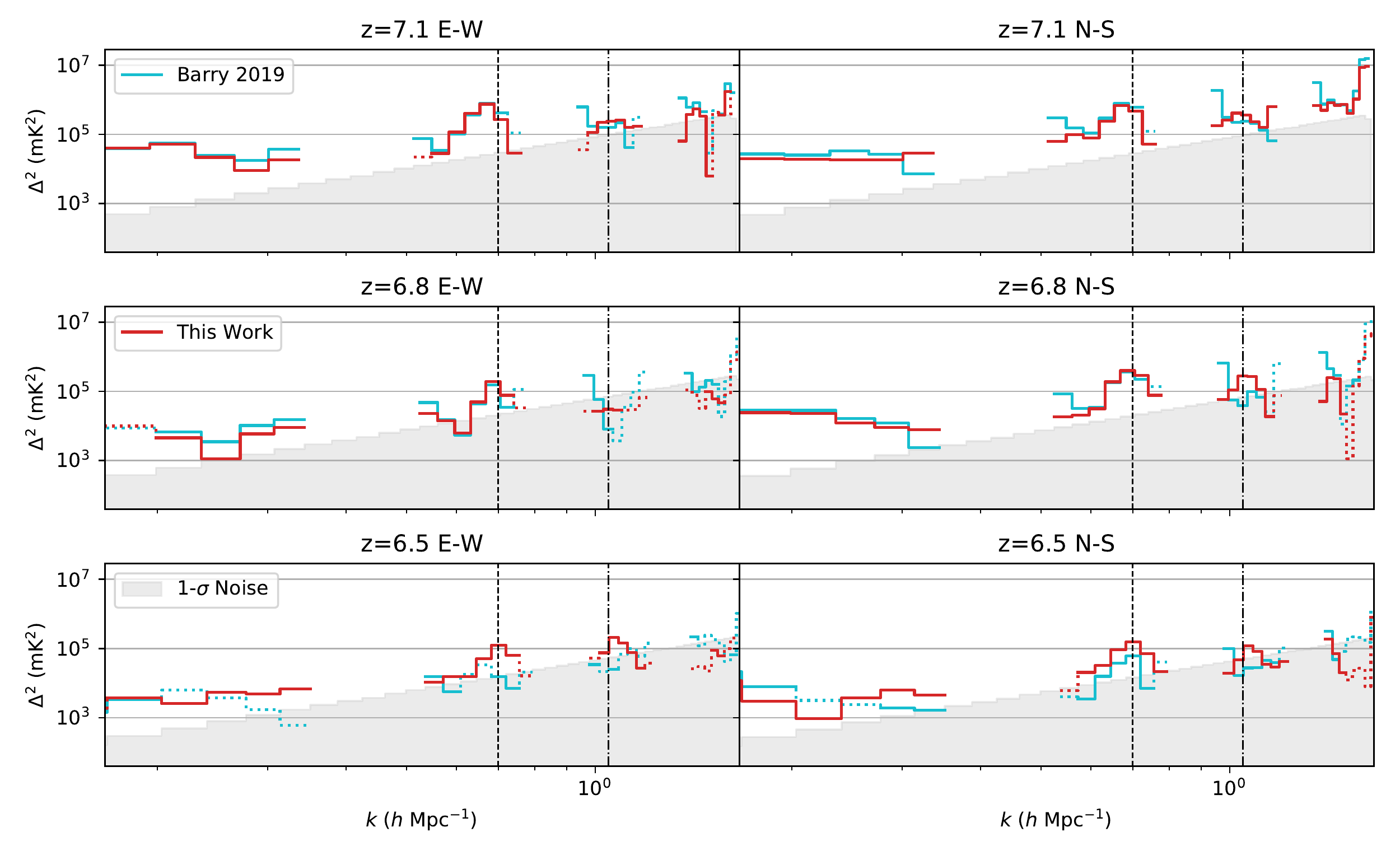}
     \caption{Contributions to power spectrum improvements at East-West polarization (left) and North-South polarization (right) in each of our three redshift bins from the auto-correlation bandpass calibration used in this work (red) and that used in \cite{barry2019newlimit} (cyan). Both techniques are compared with the bandpass calibration technique used in \cite{beardsley2016first}.  Solid lines illustrate power mitigation and dotted lines indicate an excess of power introduced by auto-correlation bandpass calibration. The shaded region quotes the 1-$\sigma$ noise level.  The vertical dashed line marks the 150 m cable reflection mode, and the vertical dashed dotted line marks the 230 m cable reflection mode.}
     \label{fig:cal_diff_bandpass}
 \end{figure*}
We can also get a better sense for the importance of our auto-correlation bandpass calibration technique using the same methodology.  Figure \ref{fig:cal_diff_bandpass} shows the differences between a power spectrum made with the \cite{beardsley2016first} bandpass calibration technique and those made with the \cite{barry2019fhd} auto-corelation bandpass calibration technique (cyan) and our auto-correlation bandpass calibration technique (red). Same as in Figure \ref{fig:cal_diff_hybrid}, solid lines show an improvement in the power spectrum, and dotted show an excess of power over the reference technique.  For both auto-correlation bandpass techniques, the changes are quite significant and represent a large improvement over the \cite{beardsley2016first} method.  Our technique generally yields improvements consistent with the \cite{barry2019fhd} technique; however, we note that our technique better mitigates contamination from modes at or near cable reflections (marked with vertical dashed and dot-dashed lines), especially in the $z=6.5$ bin.  Given that our best limits come from the E-W polarization $k = 0.59 h \mathrm{Mpc}^{-1}$ bin at $z=6.5$, Figure \ref{fig:cal_diff_bandpass} suggests that our auto-correlation bandpass technique plays a significant role in achieving these results.  Comparison with Figure \ref{fig:p1p2} also suggests a similar conclusion, where the greatest differences occur near the 150 m cable reflection mode.

\subsection{Impact of Redundancy}
For our purposes, the most significant difference between Phases I and II of the MWA is the introduction of redundant baselines into the layout.  We have already noted the improved power spectrum sensitivity of the Phase II layout and also demonstrated how it enables both the redundant calibration algorithm OMNICAL and the application of new data quality metrics.  However, there are also two negative impacts that redundancy can have on our analysis, which we describe in turn.

First, the point spread function of the instrument will be degraded compared to Phase I because of the sparser uv plane sampling.  \cite{byrne2019fundamental} demonstrate that arrays with poorer uv coverage are more susceptible to the sky-model incompleteness errors described in \cite{barry2016calibration}.  While it might be hoped that redundant calibration techniques can mitigate these errors, \cite{byrne2019fundamental} also demonstrate that the impact of redundant calibration is small in this regard.  Our findings shown in Figure \ref{fig:cal_diff_hybrid} are generally consistent with this expectation: redundant calibration itself has a small effect on our power spectrum measurements.  Because there are still 56 tiles in the pseudo-random core of the Phase II array, the uv coverage is not so degraded that sky-based calibration fails---but the increased impact of sky-model incompleteness errors is something that may have affected the analysis presented in this work.

The second way in which redundancy can negatively affect our analysis is if nominally redundant baselines are in fact non-redundant.  The two leading causes of non-redundancy are antenna position errors (i.e. the baselines are not the lengths/orientations we think they are) and antenna beam variations (i.e. the tile responses vary from tile-to-tile).  At some level, there is nothing unique to redundant arrays when dealing with these errors.  FHD's sky-based calibration uses forward-modeled visibilities to compare with the data; if the antenna positions are incorrect or beams are mis-modeled, the calibration will be affected.  We expect the antenna position errors in the Phase II array to be quite small \citep{wayth2018phase}, but antenna-to-antenna beam variations less well constrained.  \cite{line2018orbcomm} demonstrate that these beam variations are certainly present in the MWA, but their expected net effect on sky-based calibration is uncertain.  Redundant calibration is somewhat different, at least as implemented here, where perfect redundancy between redundant baselines is \emph{assumed}.  \cite{orosz2019mitigating} show how both kinds of errors can bias redundant calibration and present a modified redundant calibration algorithm that emphasizes short baselines to mitigate the bias.  Given the lack of a diffuse model used in our sky-based calibration, however, we do not believe the short baselines to be reliable calibrators, so we forgo any attempt to minimize the bias using such an approach.  Simulations using the true (i.e. measured) MWA Phase II antenna positions suggest a negligible bias in the power spectrum from this source of non-redundancy, but we cannot rule out beam variations as a source of error in our calibration.  Given the small overall effect of redundant calibration on our power spectrum measurements, however, we expect these errors are not the limiting factor in our present analysis.

\section{Conclusion}
\label{sec:conclusion}

In this work, we have presented a new analysis of MWA Phase II data using the FHD/$\varepsilon$ppsilon pipeline, including novel data quality metrics and  calibration techniques. 
The results of this work are the first MWA Phase II EoR power spectra limits at redshift 7.1, 6.8 and 6.5. We have obtained noise dominated measurements between the first and second coarse band modes, and highlight our lowest measurements at $z=6.5$. We show significant improvement in the power spectrum limits compared with \cite{beardsley2016first}, which uses an older version of the FHD-$\varepsilon$ppsilon pipeline. 
We further compare this work with the reanalysis of MWA Phase I EoR observations by \cite{barry2019newlimit}, which yields results that are generally consistent with ours using a very similar analysis. 
In addition to using the FHD/$\varepsilon$ppsilon improvements presented in \cite{barry2019fhd} and \cite{barry2019newlimit}, there are several key features that have played an important role in our analysis:

\textbf{The improved power spectrum sensitivity of MWA Phase II.} One of the motivations for the MWA Phase II design was to increase power spectrum sensitivity over MWA Phase I. Phase II has installed a large number of short baselines, especially redundant baselines, which guarantees more sampling in large scale spatial modes. The comparison between this work and \cite{barry2019newlimit} has shown that Phase II achieves lower noise levels for a comparable amount of data.
Given that many of our upper limits above $k = 0.5$ $h$ $\mathrm{Mpc^{-1}}$ are consistent with zero given the noise level, further integration with more data may improve our limit.

\textbf{New data quality metrics.} In this work, we use three principal metrics for identify anomalous data. One of them is an EoR window power metric very similar to that used in \cite{beardsley2016first}. The second is the Sky-Subtracted Incoherent Noise Spectrum (SSINS; \citealt{wilensky2019ins}) for faint RFI detection; we use our own methodology for flagging data based on SSINS which is distinct from both that proposed in \cite{wilensky2019ins} and the alternative approach used \cite{barry2019newlimit}.  A comparison of these and other different flagging techniques based on SSINS is an interesting topic for future study. The last metric that we introduce is the $\chi^2$ of redundant calibration, which is also an informative indicator for RFI detection, especially RFI from the horizon which is not detected by all antennas. The SSINS metric and redundant calibration $\chi^2$ help flag 14\%-16\% of the data in addition to \texttt{AOFlagger}. In addition, these two metrics are more computationally efficient than the window power metric.

\textbf{New calibration techniques.} Calibration is a potential limitation in any 21 cm EoR analysis. 
We have developed a bandpass calibration using auto-correlations different from \cite{ewall2016first} and \cite{barry2019fhd}.  Compared with older approaches, this technique reduces contamination in the EoR window, especially in modes below the first coarse band harmonics and near cable reflections.  The results are similar to the auto-correlation based technique used in \cite{barry2019newlimit}, but an improved elimination of 150 m cable reflection contamination in this work plays an important role in achieving our best limit at $z = 6.5$.
We also apply the hybrid calibration technique introduced in \cite{li2018comparing}, which uses both sky model based calibration and redundant calibration.  \cite{li2018comparing} demonstrated the ability of this hybrid approach to provide a small but non-negligible reduction of contamination in the EoR window; however, that work did not include our auto-correlation based bandpass calibration technique.  After adding that technique to our analysis, the improvements from redundant calibration are further reduced and they no longer appear significant.  We conclude that redundancy based calibration, as applied here, is not a major contributor to improving our final limits.
Ultimately, the GLEAM catalog provides a relatively accurate sky model for the EoR0 field, meaning that one might not expect substantial improvements from redundant calibration \citep{byrne2019fundamental}.
In future work we will investigate the performance of redundant calibration on more complicated fields where our sky model is expected to be less accurate. 

There are several future directions for the MWA EoR project to pursue for further improvements in our ability to measure to 21 cm EoR signal. A potential upgrade to the MWA correlator may get rid of the coarse band channelization, which will significantly improve our ability to calibrate a smooth bandpass. The GLEAM sky model is also a point source catalog; in future work we will include diffuse emission with a full treatment of polarization, which should improve both calibration and model subtraction.  Although our current power spectrum limit is still orders of magnitude higher than the theoretical level, analysis techniques continue to improve and our ability to reach the intrinsic sensitivity of our measurements grows with them.  The MWA has collected vastly more data than analyzed here; with continued analysis improvements, future work can place increasingly stringent limits on the EoR signal and perhaps even provide a first detection.

\acknowledgments
WL \& JCP would like to acknowledge the support from NSF grant \#1613040. WL would like to acknowledge the Galkin Foundation Fellowship. Further support for this work was provided by NSF grant \#1506024.  This scientific work makes use of the Murchison Radio-astronomy Observatory, operated by CSIRO. We acknowledge the Wajarri Yamatji people as the traditional owners of the Observatory site. Support for the operation of the MWA is provided by the Australian Government (NCRIS), under a contract to Curtin University administered by Astronomy Australia Limited. We acknowledge the Pawsey Supercomputing Centre which is supported by the Western Australian and Australian Governments. Parts of this research were supported by the Australian Research Council Centre of Excellence for All Sky Astrophysics in 3 Dimensions (ASTRO 3D), through project number CE170100013. CMT is supported by an ARC Future Fellowship under grant FT180100321. The International Centre for Radio Astronomy Research (ICRAR) is a Joint Venture of Curtin University and The University of Western Australia, funded by the Western Australian State government. The MWA Phase II upgrade project was supported by Australian Research Council LIEF grant LE160100031 and the Dunlap Institute for Astronomy and Astrophysics at the University of Toronto. This research was conducted using computation resources and services at the Center for Computation and Visualization, Brown University.

\software{cotter\, \citep{offringa2010post}, 
          $\varepsilon$ppsilon\, \citep{jacobs2016murchison,barry2019fhd},  
          FHD\, \citep{sullivan2012fast,barry2019fhd}, 
          omnical\, \citep{zheng2014miteor},
          pyuvdata\, \citep{hazelton2017pyuvdata}
}

\appendix
\section{Power spectrum Upper Limit Calculation}
\label{apdx:limcal}
As the power spectrum measurement can be negative at noise dominated bins, in the upper limit calculation, we add a prior that the power is guaranteed to be positive. We denote the true power as $x$, the prior is:
\begin{equation}
    p(x)=\systeme*{1\,\,(x\geq 0), 0\,\,(x<0)}
\end{equation}
We denote the measurement being $x^\prime$ and the variance being $\sigma^2$. The probability density function of $x^\prime$ is
\begin{equation}
    p(x^\prime|x)=\frac{1}{\sqrt{2\pi}\sigma}e^{-\frac{|x-x^\prime|^2}{\sigma^2}}
\end{equation}
The goal is to use the measurement to place an upper limit of $x$ with a confidence interval of $c$. The posterior probability is
\begin{equation}
    p(x|x^\prime)=Np(x)p(x^\prime|x)
\end{equation}
where $N$ is a normalization factor. To find $N$, we normalize $p(x|x^\prime)$:
\begin{equation}
    \begin{split}
    1&=\int_{-\infty}^{\infty} p(x|x^\prime) \mathrm{d}x \\
     &=\int_{-\infty}^{\infty} Np(x)p(x^\prime|x) \mathrm{d}x \\
     &=\int_{0}^{\infty}\frac{N}{\sqrt{2\pi}\sigma}e^{-\frac{|x-x^\prime|^2}{\sigma^2}}\mathrm{d}x \\
     &=N[\frac{1}{2}+\frac{1}{2}\erf(\frac{x^\prime}{\sqrt{2}\sigma})]
    \end{split}
\end{equation}
Therefore $N=2[1+\erf(\frac{x^\prime}{\sqrt{2}\sigma})]^{-1}$. 

To obtain the $2\sigma$ upper limit, which corresponds to $c=97.7\%$, we solve for $x_\mathrm{UL}$ where
\begin{equation}
    \int_{-\infty}^{x_\mathrm{UL}} p(x|x^\prime) \mathrm{d}x = c
\end{equation}
The solution to $x_\mathrm{UL}$ is
\begin{equation}
    x_\mathrm{UL}=x^\prime+\sqrt{2}\sigma\erf^{-1}(c-(1-c)\erf(\frac{x^\prime}{\sqrt{2}\sigma}))
\end{equation}

\section{All calculated EoR power spectra upper limits}

To aid future studies in comparing with our results, we present tables containing all our $2-\sigma$ power spectrum upper limits from both polarizations, where $\Delta_\mathrm{UL}$ denotes the $2-\sigma$ upper limit, and $\sigma$ is the noise level.

\begin{table}
\centering
\caption{Power spectrum upper limit for East-West Polarization}
\begin{tabular}{|| c c c | c c c | c c c ||}
\hline
 & $z=7.1$ & & & $z=6.8$ & & & $z=6.5$ & \\
\hline
$k$ ($h$ $\mathrm{Mpc^{-1}}$) & $\Delta_\mathrm{UL}^2$ ($\mathrm{mK^2}$) & $\sigma$ ($\mathrm{mK^2}$) & $k$ ($h$ $\mathrm{Mpc^{-1}}$) & $\Delta_\mathrm{UL}^2$ ($\mathrm{mK^2}$) & $\sigma$ ($\mathrm{mK^2}$) & $k$ ($h$ $\mathrm{Mpc^{-1}}$) & $\Delta_\mathrm{UL}^2$ ($\mathrm{mK^2}$) & $\sigma$ ($\mathrm{mK^2}$) \\
\hline
\hline
$0.177$ & $5.37\times 10^4$ & $1.33\times 10^2$ & $0.181$ & $2.57\times 10^4$ & $1.04\times 10^2$ & $0.185$ & $1.20\times 10^4$ & $8.40\times 10^1$\\ 
 \hline
$0.212$ & $1.44\times 10^4$ & $2.04\times 10^2$ & $0.217$ & $9.41\times 10^3$ & $1.58\times 10^2$ & $0.222$ & $1.06\times 10^4$ & $1.27\times 10^2$\\ 
 \hline
$0.248$ & $1.58\times 10^4$ & $3.22\times 10^2$ & $0.253$ & $9.74\times 10^3$ & $2.49\times 10^2$ & $0.258$ & $1.06\times 10^4$ & $2.01\times 10^2$\\ 
 \hline
$0.283$ & $1.96\times 10^4$ & $4.78\times 10^2$ & $0.289$ & $1.11\times 10^4$ & $3.69\times 10^2$ & $0.295$ & $8.53\times 10^3$ & $2.96\times 10^2$\\ 
 \hline
$0.318$ & $3.06\times 10^4$ & $6.77\times 10^2$ & $0.325$ & $1.50\times 10^4$ & $5.22\times 10^2$ & $0.332$ & $1.04\times 10^4$ & $4.18\times 10^2$\\ 
 \hline
$0.531$ & $7.95\times 10^4$ & $3.14\times 10^3$ & $0.542$ & $5.18\times 10^4$ & $2.42\times 10^3$ & $0.554$ & $8.76\times 10^3$ & $1.94\times 10^3$\\ 
 \hline
$0.566$ & $1.30\times 10^4$ & $3.81\times 10^3$ & $0.579$ & $2.50\times 10^4$ & $2.93\times 10^3$ & $0.591$ & $2.39\times 10^3$ & $2.35\times 10^3$\\ 
 \hline
$0.601$ & $2.60\times 10^4$ & $4.56\times 10^3$ & $0.615$ & $2.68\times 10^4$ & $3.51\times 10^3$ & $0.628$ & $6.05\times 10^3$ & $2.81\times 10^3$\\ 
 \hline
$0.637$ & $3.48\times 10^4$ & $5.37\times 10^3$ & $0.651$ & $2.81\times 10^4$ & $4.14\times 10^3$ & $0.665$ & $2.07\times 10^4$ & $3.32\times 10^3$\\ 
 \hline
$0.672$ & $1.93\times 10^4$ & $6.36\times 10^3$ & $0.687$ & $3.95\times 10^4$ & $4.91\times 10^3$ & $0.702$ & $3.81\times 10^4$ & $3.93\times 10^3$\\ 
 \hline
$0.708$ & $4.78\times 10^4$ & $7.43\times 10^3$ & $0.723$ & $8.04\times 10^4$ & $5.73\times 10^3$ & $0.739$ & $5.53\times 10^4$ & $4.59\times 10^3$\\ 
 \hline
$0.743$ & $1.36\times 10^5$ & $8.61\times 10^3$ & $0.759$ & $1.28\times 10^5$ & $6.63\times 10^3$ & $0.775$ & $9.44\times 10^4$ & $5.31\times 10^3$\\ 
 \hline
$0.955$ & $6.56\times 10^5$ & $1.83\times 10^4$ & $0.976$ & $3.82\times 10^5$ & $1.41\times 10^4$ & $0.997$ & $1.78\times 10^5$ & $1.13\times 10^4$\\ 
 \hline
$0.991$ & $1.29\times 10^5$ & $2.04\times 10^4$ & $1.012$ & $1.86\times 10^5$ & $1.57\times 10^4$ & $1.034$ & $1.09\times 10^5$ & $1.26\times 10^4$\\ 
 \hline
$1.026$ & $1.45\times 10^5$ & $2.26\times 10^4$ & $1.049$ & $1.98\times 10^5$ & $1.74\times 10^4$ & $1.071$ & $4.75\times 10^4$ & $1.40\times 10^4$\\ 
 \hline
$1.061$ & $2.37\times 10^5$ & $2.49\times 10^4$ & $1.085$ & $1.36\times 10^5$ & $1.92\times 10^4$ & $1.108$ & $1.87\times 10^4$ & $1.54\times 10^4$\\ 
 \hline
$1.097$ & $2.59\times 10^5$ & $2.76\times 10^4$ & $1.121$ & $9.92\times 10^4$ & $2.13\times 10^4$ & $1.145$ & $1.23\times 10^4$ & $1.71\times 10^4$\\ 
 \hline
$1.132$ & $2.17\times 10^5$ & $3.04\times 10^4$ & $1.157$ & $1.71\times 10^5$ & $2.35\times 10^4$ & $1.182$ & $4.52\times 10^4$ & $1.88\times 10^4$\\ 
 \hline
$1.167$ & $8.17\times 10^5$ & $3.34\times 10^4$ & $1.193$ & $3.95\times 10^5$ & $2.57\times 10^4$ & $1.219$ & $2.39\times 10^5$ & $2.06\times 10^4$\\ 
 \hline
$1.380$ & $1.96\times 10^6$ & $5.51\times 10^4$ & $1.410$ & $1.12\times 10^6$ & $4.25\times 10^4$ & $1.440$ & $5.10\times 10^5$ & $3.40\times 10^4$\\ 
 \hline
$1.415$ & $3.07\times 10^5$ & $5.95\times 10^4$ & $1.446$ & $4.16\times 10^5$ & $4.59\times 10^4$ & $1.477$ & $2.42\times 10^5$ & $3.67\times 10^4$\\ 
 \hline
$1.450$ & $7.85\times 10^4$ & $6.36\times 10^4$ & $1.482$ & $3.52\times 10^5$ & $4.90\times 10^4$ & $1.514$ & $2.06\times 10^5$ & $3.92\times 10^4$\\ 
 \hline
$1.486$ & $7.76\times 10^4$ & $6.62\times 10^4$ & $1.519$ & $2.08\times 10^5$ & $5.10\times 10^4$ & $1.551$ & $2.39\times 10^5$ & $4.08\times 10^4$\\ 
 \hline
$1.521$ & $7.05\times 10^4$ & $7.33\times 10^4$ & $1.555$ & $3.50\times 10^5$ & $5.65\times 10^4$ & $1.588$ & $3.09\times 10^5$ & $4.53\times 10^4$\\ 
 \hline
$1.557$ & $8.48\times 10^5$ & $7.92\times 10^4$ & $1.591$ & $6.46\times 10^5$ & $6.10\times 10^4$ & $1.625$ & $3.64\times 10^5$ & $4.89\times 10^4$\\ 
 \hline
$1.592$ & $4.89\times 10^6$ & $8.47\times 10^4$ & $1.627$ & $1.71\times 10^6$ & $6.53\times 10^4$ & $1.662$ & $8.54\times 10^5$ & $5.23\times 10^4$\\ 
 \hline
\end{tabular}
 \label{tb:bestlimx}
\end{table}

\begin{table}
\centering
\caption{Power spectrum upper limit for North-South Polarization}
\begin{tabular}{|| c c c | c c c | c c c ||}
\hline
 & $z=7.1$ & & & $z=6.8$ & & & $z=6.5$ & \\
\hline
$k$ ($h$ $\mathrm{Mpc^{-1}}$) & $\Delta_\mathrm{UL}^2$ ($\mathrm{mK^2}$) & $\sigma$ ($\mathrm{mK^2}$) & $k$ ($h$ $\mathrm{Mpc^{-1}}$) & $\Delta_\mathrm{UL}^2$ ($\mathrm{mK^2}$) & $\sigma$ ($\mathrm{mK^2}$) & $k$ ($h$ $\mathrm{Mpc^{-1}}$) & $\Delta_\mathrm{UL}^2$ ($\mathrm{mK^2}$) & $\sigma$ ($\mathrm{mK^2}$) \\
\hline
\hline
$0.177$ & $4.53\times 10^4$ & $1.14\times 10^2$ & $0.181$ & $3.25\times 10^4$ & $8.75\times 10^1$ & $0.185$ & $2.51\times 10^4$ & $6.97\times 10^1$\\ 
 \hline
$0.212$ & $1.19\times 10^4$ & $1.75\times 10^2$ & $0.217$ & $1.52\times 10^4$ & $1.33\times 10^2$ & $0.222$ & $1.18\times 10^4$ & $1.06\times 10^2$\\ 
 \hline
$0.248$ & $9.28\times 10^3$ & $2.76\times 10^2$ & $0.253$ & $1.05\times 10^4$ & $2.10\times 10^2$ & $0.258$ & $7.82\times 10^3$ & $1.66\times 10^2$\\ 
 \hline
$0.283$ & $9.93\times 10^3$ & $4.09\times 10^2$ & $0.289$ & $5.59\times 10^3$ & $3.10\times 10^2$ & $0.295$ & $6.24\times 10^3$ & $2.45\times 10^2$\\ 
 \hline
$0.318$ & $1.50\times 10^4$ & $5.80\times 10^2$ & $0.325$ & $6.98\times 10^3$ & $4.40\times 10^2$ & $0.332$ & $1.01\times 10^4$ & $3.46\times 10^2$\\ 
 \hline
$0.531$ & $2.48\times 10^4$ & $2.68\times 10^3$ & $0.542$ & $2.71\times 10^4$ & $2.03\times 10^3$ & $0.554$ & $2.20\times 10^4$ & $1.60\times 10^3$\\ 
 \hline
$0.566$ & $1.24\times 10^4$ & $3.26\times 10^3$ & $0.579$ & $2.05\times 10^4$ & $2.47\times 10^3$ & $0.591$ & $1.03\times 10^4$ & $1.95\times 10^3$\\ 
 \hline
$0.601$ & $1.54\times 10^4$ & $3.90\times 10^3$ & $0.615$ & $1.19\times 10^4$ & $2.96\times 10^3$ & $0.628$ & $5.47\times 10^3$ & $2.33\times 10^3$\\ 
 \hline
$0.637$ & $4.82\times 10^4$ & $4.60\times 10^3$ & $0.651$ & $1.25\times 10^4$ & $3.49\times 10^3$ & $0.665$ & $8.29\times 10^3$ & $2.75\times 10^3$\\ 
 \hline
$0.672$ & $7.06\times 10^4$ & $5.45\times 10^3$ & $0.687$ & $1.21\times 10^4$ & $4.13\times 10^3$ & $0.702$ & $1.53\times 10^4$ & $3.25\times 10^3$\\ 
 \hline
$0.708$ & $5.39\times 10^4$ & $6.36\times 10^3$ & $0.723$ & $3.31\times 10^4$ & $4.82\times 10^3$ & $0.739$ & $1.55\times 10^4$ & $3.80\times 10^3$\\ 
 \hline
$0.743$ & $1.02\times 10^5$ & $7.36\times 10^3$ & $0.759$ & $1.00\times 10^5$ & $5.58\times 10^3$ & $0.775$ & $4.79\times 10^4$ & $4.40\times 10^3$\\ 
 \hline
$0.955$ & $2.79\times 10^5$ & $1.57\times 10^4$ & $0.976$ & $2.63\times 10^5$ & $1.19\times 10^4$ & $0.997$ & $2.70\times 10^5$ & $9.35\times 10^3$\\ 
 \hline
$0.991$ & $9.83\times 10^4$ & $1.75\times 10^4$ & $1.012$ & $1.13\times 10^5$ & $1.32\times 10^4$ & $1.034$ & $6.14\times 10^4$ & $1.04\times 10^4$\\ 
 \hline
$1.026$ & $1.77\times 10^5$ & $1.94\times 10^4$ & $1.049$ & $9.29\times 10^4$ & $1.47\times 10^4$ & $1.071$ & $1.06\times 10^5$ & $1.16\times 10^4$\\ 
 \hline
$1.061$ & $2.13\times 10^5$ & $2.13\times 10^4$ & $1.085$ & $1.50\times 10^5$ & $1.61\times 10^4$ & $1.108$ & $6.43\times 10^4$ & $1.27\times 10^4$\\ 
 \hline
$1.097$ & $2.04\times 10^5$ & $2.37\times 10^4$ & $1.121$ & $1.38\times 10^5$ & $1.79\times 10^4$ & $1.145$ & $3.13\times 10^4$ & $1.41\times 10^4$\\ 
 \hline
$1.132$ & $2.72\times 10^5$ & $2.61\times 10^4$ & $1.157$ & $8.73\times 10^4$ & $1.98\times 10^4$ & $1.182$ & $9.47\times 10^4$ & $1.56\times 10^4$\\ 
 \hline
$1.167$ & $6.06\times 10^5$ & $2.86\times 10^4$ & $1.193$ & $2.67\times 10^5$ & $2.17\times 10^4$ & $1.219$ & $2.56\times 10^5$ & $1.71\times 10^4$\\ 
 \hline
$1.380$ & $6.36\times 10^5$ & $4.72\times 10^4$ & $1.410$ & $2.34\times 10^5$ & $3.58\times 10^4$ & $1.440$ & $5.17\times 10^5$ & $2.82\times 10^4$\\ 
 \hline
$1.415$ & $3.42\times 10^5$ & $5.09\times 10^4$ & $1.446$ & $5.92\times 10^4$ & $3.86\times 10^4$ & $1.477$ & $1.04\times 10^5$ & $3.04\times 10^4$\\ 
 \hline
$1.450$ & $1.15\times 10^5$ & $5.44\times 10^4$ & $1.482$ & $4.43\times 10^4$ & $4.12\times 10^4$ & $1.514$ & $1.83\times 10^5$ & $3.25\times 10^4$\\ 
 \hline
$1.486$ & $3.95\times 10^4$ & $5.66\times 10^4$ & $1.519$ & $6.19\times 10^4$ & $4.29\times 10^4$ & $1.551$ & $6.45\times 10^4$ & $3.38\times 10^4$\\ 
 \hline
$1.521$ & $4.05\times 10^4$ & $6.28\times 10^4$ & $1.555$ & $3.26\times 10^5$ & $4.76\times 10^4$ & $1.588$ & $3.65\times 10^4$ & $3.75\times 10^4$\\ 
 \hline
$1.557$ & $8.31\times 10^4$ & $6.77\times 10^4$ & $1.591$ & $4.07\times 10^5$ & $5.14\times 10^4$ & $1.625$ & $6.17\times 10^4$ & $4.05\times 10^4$\\ 
 \hline
$1.592$ & $1.32\times 10^6$ & $7.25\times 10^4$ & $1.627$ & $1.19\times 10^6$ & $5.49\times 10^4$ & $1.662$ & $1.04\times 10^6$ & $4.33\times 10^4$\\ 
 \hline
\end{tabular}
 \label{tb:bestlimy}
\end{table}


\bibliographystyle{aasjournal}
\bibliography{library}

\end{document}